\documentclass[aps,prl,reprint,twocolumn,superscriptaddress,floatfix,nofootinbib,longbibliography]{revtex4-2}
\usepackage{graphicx,amsmath,amsfonts,amssymb,amsthm,xr}

\usepackage{epsfig,amsmath,amssymb,color,dsfont,upgreek,physics}
\usepackage{blkarray,bigstrut}
\usepackage{mathrsfs}
\usepackage{mathtools}
\usepackage{bbold}

\usepackage{float}
\usepackage[caption=false]{subfig}

\usepackage[bookmarks=true,colorlinks,linkcolor=OrangeRed,urlcolor=NavyBlue,citecolor=RoyalBlue]{hyperref}
\usepackage[dvipsnames]{xcolor}
\usepackage{orcidlink}
\usepackage[utf8]{inputenc}

\usepackage[normalem]{ulem}

\definecolor{sgrcol}{RGB}{100,75,255}

\begin{document}

\preprint{APS/123-QED}

\title{Repulsively Bound Hadrons in a $\mathbb{Z}_2$ Lattice Gauge Theory}

\author{Sayak Guha Roy${}^{\orcidlink{0000-0001-5816-6079}}$}
\email{sg161@rice.edu}
\affiliation{Department of Physics and Astronomy, Rice University, Houston, TX 77005 and Smalley-Curl Institute, Rice University, Houston, TX 77005}

\author{Vaibhav Sharma${}^{\orcidlink{0000-0003-3151-2968}}$}
\affiliation{Department of Physics and Astronomy, Rice University, Houston, TX 77005 and Smalley-Curl Institute, Rice University, Houston, TX 77005}

\author{Kaidi Xu${}^{\orcidlink{0000-0003-2184-0829}}$}
\affiliation{Max Planck Institute of Quantum Optics, 85748 Garching, Germany}
\affiliation{Munich Center for Quantum Science and Technology (MCQST), 80799 Munich, Germany}

\author{Umberto Borla${}^{\orcidlink{0000-0002-4224-5335}}$}
\affiliation{Max Planck Institute of Quantum Optics, 85748 Garching, Germany}
\affiliation{Munich Center for Quantum Science and Technology (MCQST), 80799 Munich, Germany}

\author{Jad C.~Halimeh${}^{\orcidlink{0000-0002-0659-7990}}$}
\affiliation{Department of Physics and Arnold Sommerfeld Center for Theoretical Physics (ASC), Ludwig Maximilian University of Munich, 80333 Munich, Germany}
\affiliation{Max Planck Institute of Quantum Optics, 85748 Garching, Germany}
\affiliation{Munich Center for Quantum Science and Technology (MCQST), 80799 Munich, Germany}
\affiliation{Department of Physics, College of Science, Kyung Hee University, Seoul 02447, Republic of Korea}

\author{Kaden R.~A.~Hazzard${}^{\orcidlink{0000-0003-2894-7274}}$}
\affiliation{Department of Physics and Astronomy, Rice University, Houston, TX 77005 and Smalley-Curl Institute, Rice University, Houston, TX 77005}

\date{\today}

\begin{abstract}
 
The $\mathbb{Z}_2$ lattice gauge theory is a paradigmatic model that exhibits gauge-field-mediated-confinement of pairs of particles into mesons, drawing connections to quantum chromodynamics. In the absence of any additional attractive interactions between particles, mesons are not known to bind in this model. Here, we show that resonant pair-production terms give rise to two separate mechanisms to form stable ``hadron'' bound states of two mesons: either induced by an effective attractive interaction, or a new dynamical binding mechanism induced by an effective repulsion. The repulsively bound hadron is a high-energy state stabilized by being energetically separated from  the two-meson  continuum through quantum fluctuations of the gauge fields. We study the dynamical formation of this bound state starting from local excitations. We use matrix product state techniques based on the time-evolving block decimation algorithm to perform our numerical simulations and analyze the effect of model parameters on hadron formation. Furthermore, we derive an effective model that explains its formation. Our findings are amenable to experimental observation on modern quantum hardware such as superconducting qubits,  trapped ions, and Rydberg atom arrays.

\end{abstract}

\maketitle

\textbf{\textit{Introduction.---}}Lattice gauge theories (LGTs) \cite{Rothe_book}, originally conceived to investigate the problem of quark confinement in quantum chromodynamics (QCD) \cite{Wilson1974}, have since become a powerful framework spanning a multitude of fields beyond high-energy physics (HEP) \cite{halimeh2025rmp}. In condensed matter, they serve as models for emergent gauge structures in quantum spin liquids and frustrated magnets, and they have also been invoked in certain theoretical approaches to high-temperature superconductivity \cite{Wegner1971,Kogut_review,wen2004quantum,Savary2016,Calzetta_book}. In quantum many-body physics, they have become a paradigm of nonergodic phenomena such as quantum many-body scarring \cite{Surace2020,Desaules2022weak,Desaules2022prominent,aramthottil2022scar,osborne_quantum_2024,budde_quantum_2024}, Hilbert-space fragmentation \cite{Desaules2024ergodicitybreaking,ciavarella2025generichilbertspacefragmentation,jeyaretnam2025hilbertspacefragmentationorigin}, and disorder-free localization \cite{Smith2017,Brenes2018,smith2017absence,karpov2021disorder,Chakraborty2022,osborne_disorder-free_2023}.

Whereas Monte Carlo techniques based on the Euclidean path integral formulation have enabled precision computations of decay constants, hadron masses, and thermodynamic properties of QCD \cite{Creutz1980MC,QCD_review,Gattringer_book,Montvay_book}, they face the notorious sign problem at high matter densities or out of equilibrium \cite{deforcrand2010simulatingqcdfinitedensity,Troyer2005computational}. Here, tensor network methods \cite{Uli_review,Paeckel_review,Montangero_book,Orus2019,Silvi_LectureNotes} offer powerful first-principles study of the out-of-equilibrium dynamics of LGTs \cite{Banuls2018,Banuls2020}, especially in low dimensions. At the same time, lattice gauge theories serve as a complementary platform for benchmarking and guiding the rapidly growing effort to quantum simulate high-energy physics (HEP) on quantum hardware~\cite{Byrnes2006simulating,Dalmonte_review, Zohar_review, aidelsburger2021cold, Zohar_NewReview, klco2021standard,Bauer_ShortReview, Bauer_review, dimeglio2023quantum, Cheng_review,Halimeh_BriefReview, Halimeh_review, Cohen:2021imf,Lee:2024jnt,Turro:2024pxu,bauer2025efficientusequantumcomputers}. In fact, recent quantum simulations of LGTs\cite{Martinez2016,Klco2018,Goerg2019,Schweizer2019,Mil2020,Yang2020,Wang2021,Su2022,Zhou2022,Wang2023,Zhang2023,Ciavarella2024quantum,Ciavarella:2024lsp,Farrell:2023fgd,Farrell:2024fit,zhu2024probingfalsevacuumdecay,Ciavarella:2021nmj,Ciavarella:2023mfc,Ciavarella:2021lel,Gustafson:2023kvd,Gustafson:2024kym,Lamm:2024jnl,Farrell:2022wyt,Farrell:2022vyh,de_observation_2024,liu2024stringbreakingmechanismlattice,Li:2024lrl,Zemlevskiy:2024vxt,Lewis:2019wfx,Atas:2021ext,ARahman:2022tkr,Atas:2022dqm,Mendicelli:2022ntz,Kavaki:2024ijd,Than:2024zaj,Mildenberger2025,Angelides2025first,cochran2024visualizingdynamicschargesstrings,alexandrou2025realizingstringbreakingdynamics,gyawali2024observationdisorderfreelocalizationefficient,gonzalezcuadra2024observationstringbreaking2,crippa2024analysisconfinementstring2,schuhmacher2025observationhadronscatteringlattice,davoudi2025quantumcomputationhadronscattering,cobos2025realtimedynamics21dgauge,saner2025realtimeobservationaharonovbohminterference,xiang2025realtimescatteringfreezeoutdynamics,wang2025observationinelasticmesonscattering,mark2025observationballisticplasmamemory,darbha2025probingemergentprethermaldynamics} have led to increased interest in analyzing LGTs with tensor network methods to guide future experiments \cite{pichler_real-time_2016,chanda_confinement_2020,rigobello_entanglement_2021,van_damme_dynamical_2022,su_cold-atom_2024,calajo_quantum_2025,cataldi_disorder-free_2025,belyansky_high-energy_2024}. 

\begin{figure}[t!]
    \centering
    \includegraphics[width=8cm]{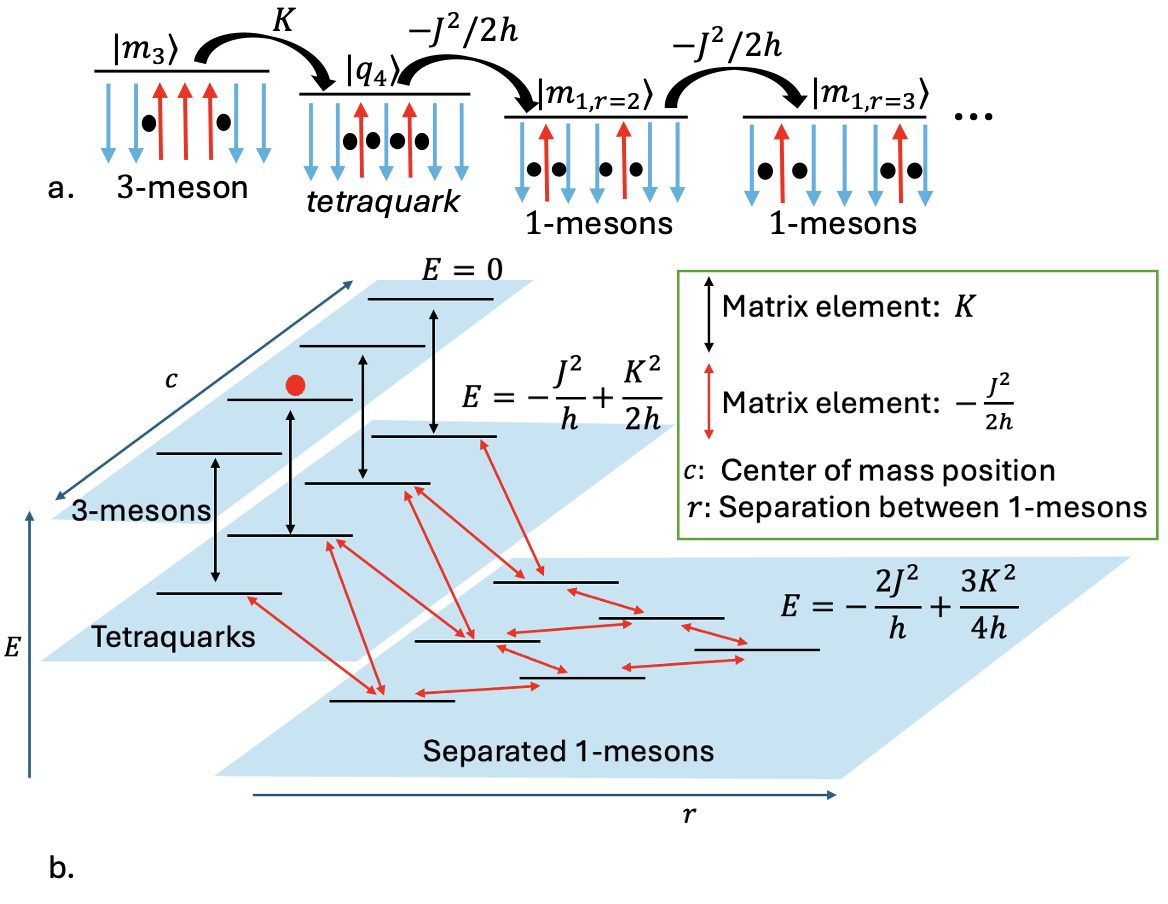}
    \caption{{\bf a}. Examples of spin and matter configurations for the $\mathbb{Z}_2$ lattice gauge theory defined by the Hamiltonian Eq.~\eqref{eq:H_fsp}. Circles correspond to matter particles, which are present when there is a domain wall in gauge spins. We call two matter particles separated by $n$ spins an $n$-meson and four particles next to each other a \emph{tetraquark} $\ket{q_4}$ state. {\bf b}. The effective model in the $h,m\gg J,K$ limit, showing the relevant $3$-meson and \emph{tetraquark} states occupying different center-of-mass positions $c$ and separated $1$-meson states with different relative separations $r$, along with the corresponding diagonal and off-diagonal matrix elements connecting them.}
    \label{fig:phenmlgy}
\end{figure}

Of particular interest is the $\mathbb{Z}_2$ LGT \cite{Wegner1971,Fradkin1979}, which has recently been studied in several experiments in one \cite{de_observation_2024,Mildenberger2025,wang2025observationinelasticmesonscattering} and two \cite{cochran2024visualizingdynamicschargesstrings,gyawali2024observationdisorderfreelocalizationefficient} spatial dimensions. The $\mathbb{Z}_2$ LGT with dynamical matter has been successful in providing insights into phenomena such as confinement \cite{Borla2020confined,GonzalezCuadra2020robust,kebric2021confinement,Homeier2023,Fromm2024,Kebric2024FiniteTconfinement,Linsel2024}, the Higgs mechanism \cite{Assaad2016simple,Gazit2018,Koening2020soluble,Somoza2021self-dual,Borla2024deconfined,xu2024critical}, exotic phase transitions \cite{Borla2020,verresen2024}, string breaking \cite{borla2025stringbreaking21dmathbbz2,xu2025tensornetworkstudyrougheningtransition,xu2025stringbreakingdynamicsglueball,tian2025roleplaquettetermgenuine}, and scattering \cite{Surace2021scattering,davoudi2025quantumcomputationhadronscattering}. This body of work motivates the exploration of excitations in this model that can give a fundamental understanding of its properties and also guide quantum simulation experiments of HEP-relevant phenomena.

In this paper, we simulate the dynamics of a $\mathbb{Z}_2$ LGT in $1+1$D and observe the phenomena of repulsive bound states of two mesons. These are exotic physical states that arise as a result of repulsion as opposed to attraction. They have been observed to arise in lattice models where there is an absence of dissipation to a lower-energy continuum \cite{Winkler_2006}. They have also been observed in spin chains with staggered fields \cite{Rep_bound1}. In our work, we show that even in the presence of a low-energy continuum of two-meson states, if we initialize our system in simple higher-energy states with average energy above the continuum, a finite fraction of the excitation remains stable, never dissociating. We argue that this is due to formation of a repulsively bound state stabilized by quantum fluctuations of gauge fields. These states have not been observed in gauge theories, and in general are not expected to be stable in continuum gauge theories since the continuum is unbounded from above. We note that these bound states are different from the ones observed due to additional long-range interactions among particles~\cite{hadron1,hadron2}. The $\mathbb{Z}_2$ lattice gauge theory that we study can be realized in experiments where one can, for the first time, observe repulsive bound states in a gauge theory.

\textbf{\textit{Model.---}}We consider a particle non-conserving $\mathbb{Z}_2$ lattice gauge theory in which hard-core bosons are coupled to spin-1/2 gauge fields, governed by the Hamiltonian
\begin{align}
    \hat{\mathcal{H}} = &-J\sum_{i} \big(\hat{b}_{i+1}^{\dagger} \hat{\sigma}^x_{i,i+1}\hat{b}_i + \text{H.c.}\big) + h \sum_i \hat{\sigma}^z_{i,i+1} \nonumber\\
    &-K \sum_i \big(\hat{b}_{i+1}^{\dagger} \hat{\sigma}^x_{i,i+1} \hat{b}_i^{\dagger} +\text{H.c.}\big) +m\sum_i \hat{b}^{\dagger}_i \hat{b}_i \,.
    \label{eq:H_fsp}
\end{align}
Here, $\hat{b}_i$ ($\hat{b}_i^{\dagger}$) annihilates (creates) a boson at site $i$ and $\hat{\sigma}_{i,i+1}^{\mu}$ ($\mu = x,y,z$) is the Pauli spin-1/2 operator acting on the link between sites $i$ and $i+1$. The parameter $J$ controls single-particle hopping accompanied by a gauge spin-flip on the traversed link, $h$ is the electric field term, $K$ governs boson pair creation and annihilation processes, and $m$ denotes the particle rest mass.

This Hamiltonian respects a local $\mathbb{Z}_2$ gauge invariance generated by the local operators $\hat{G}_i = \hat{\sigma}^z_{i-1,i}(-1)^{\hat{b}_i^{\dagger}\hat{b}_i}\hat{\sigma}^z_{i,i+1}$ \cite{Meson_dynamics_Vaibhav}. $\hat{G}_i$ transforms the Hamiltonian operators as $\hat{G}_i \hat{b}_i \hat{G}_i^{\dagger} = -\hat{b}_i$, $\hat{G}_i \hat{b}_i^{\dagger}\hat{G}_i^{\dagger} = -\hat{b}_i^{\dagger}$, $\hat{G}_i \hat{\sigma}^x_{i,i+1}\hat{G}_i^{\dagger} = -\hat{\sigma}^x_{i,i+1}$ and $\hat{G}_i \hat{\sigma}^z_{i,i+1}\hat{G}_i^{\dagger} = \hat{\sigma}^z_{i,i+1}$. Consequently, $[\hat{\mathcal{H}},\hat{G}_i]=0$, ensuring $\mathbb{Z}_2$ gauge invariance. A specific configuration of eigenvalues of $\hat{G}_i$ enforces a Gauss law constraint that restricts the possible matter and spin configurations. In this work, we consider the sector with $\hat{G}_i=+1$ $\forall$ $i$ as shown in Fig.~\ref{fig:phenmlgy}(a), where a gauge spin domain wall corresponds to a matter particle. The physical interpretation of choosing other $\hat{G}_i$ sectors has been explored in \cite{Sharma_2024}.

Using the gauge constraint, the matter degrees of freedom can be integrated out, yielding a spin-only Hamiltonian \cite{Sharma_2024} given by

\begin{align}
    \hat{\mathcal{H}} = &-\frac{J+K}{2}\sum_{i'} \hat{\tau}_{i'}^x + \frac{J-K}{2}\sum_i\hat{\tau}_{i-1}^z\hat{\tau}_i^x\hat{\tau}_{i+1}^z \nonumber\\
    & +h\sum_i \hat{\tau}^z_{i} + \frac{m}{2} \sum_i (1-\hat{\tau}^z_{i-1}\hat{\tau}^z_{i})\,,
    \label{eq:H_spin}
\end{align}
where the sites correspond to the links of the chain in Eq.~\eqref{eq:H_fsp} and $\hat{\tau}_i^{\mu}$ are the Pauli spin-1/2 operators encoding the gauge fields. The sum over $i'$ corresponds to all the spins except the first and the last. Domain walls in the spin chain correspond to a matter particle. The $J$ term causes domain wall hopping while the $K$ term creates or destroys two adjacent domain walls with an energy cost encoded by the $m$ term. The $h$ term corresponds to the electric field; large $h$ favors closely spaced domain walls leading to confinement of particles into mesons. Later we will simulate the dynamics of the spin Hamiltonian in Eq.~\eqref{eq:H_spin} using a tensor network method based on Matrix Product States \cite{Vidal2004} and develop an effective tight-binding model to understand the dynamics. Setting $K=0$ or $m \to \infty$ reproduces the meson dynamics seen in \cite{Meson_dynamics_Vaibhav}.

\textbf{\textit{Hadron Binding Phenomenology.---}}In this section, we qualitatively outline the dynamical formation and stability of four-particle hadronic states from two mechanisms: (i) repulsive binding above the lower-energy mesonic continuum, and (ii) attractive binding induced by strong particle number fluctuations that lower the state's energy. For clarity, we adopt the nomenclature shown in Fig.~\ref{fig:phenmlgy}(a) where a pair of domain walls separated by $(n-1)$ spins is called an $n$-meson (e.g., a $1$-meson contains adjacent domain walls). We call four adjacent domain walls a \emph{tetraquark} and treat it distinctly from two $1$-mesons that are nonadjacent. Due to confinement caused by the $h$ term, particles are already bound into mesons. We now show how the \emph{tetraquark} state also remains bound.

The simplest setting to present our \emph{tetraquark} bound state argument is by initializing our spin chain  with a $3$-meson configuration [Fig.~\ref{fig:phenmlgy}(a)].
In the limit where $h,m\gg J,K$ and $h=m$, this state is comparable in energy to the \emph{teraquark} state. All the relevant states in the resulting dynamics are shown in Fig.~\ref{fig:phenmlgy}(b). These are the $3$-meson states ($\ket{m_3}$), the tetraquark states ($\ket{q_4}$) and a continuum of nonadjacent $1$-mesons ($\ket{m_{1,r=2}}$ denotes a continuum state with separation $r$ between the $1$-mesons).

\begin{figure}
    \centering
    \includegraphics[width=8.6cm]{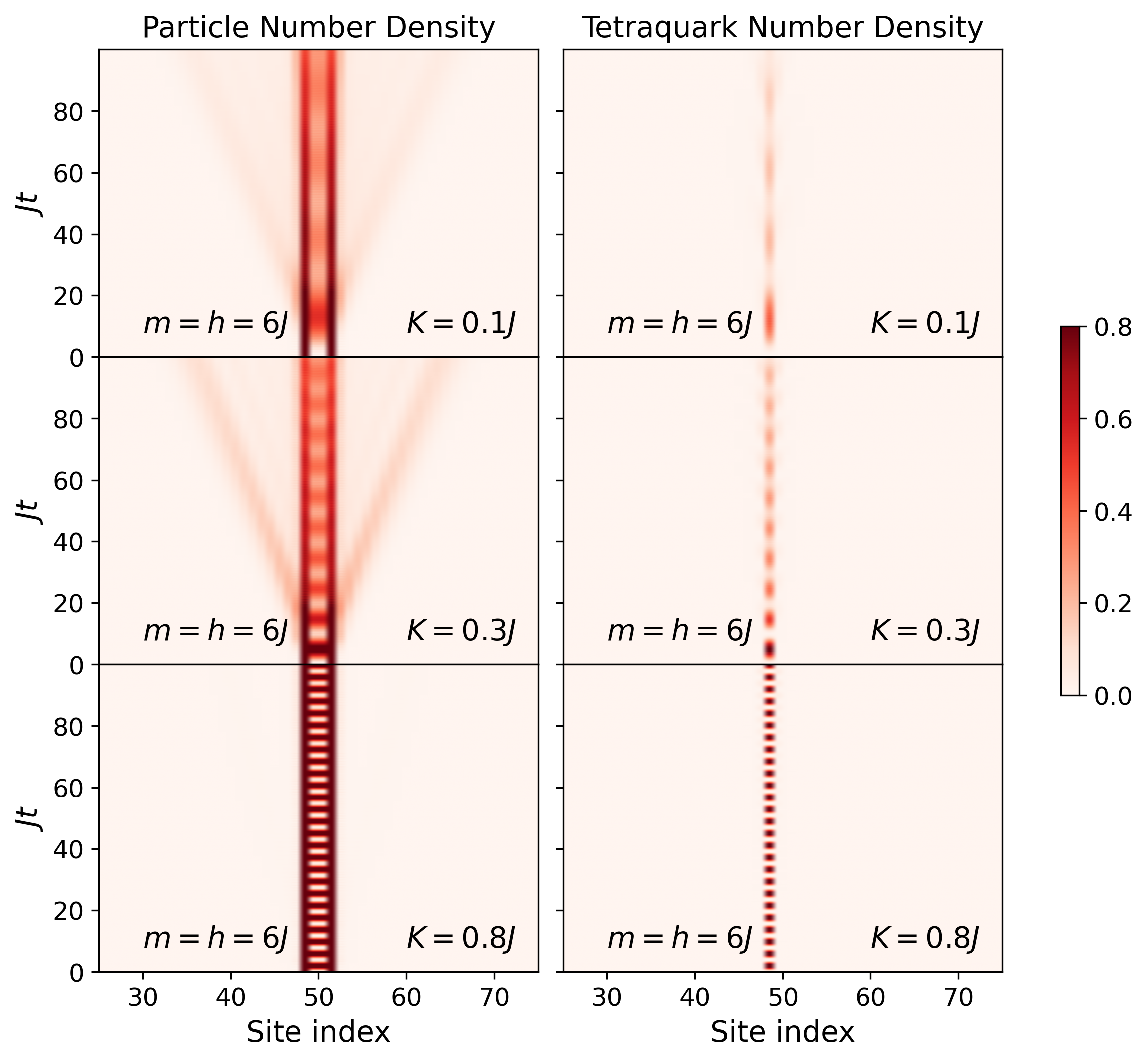}
    \caption{Density plots of the local particle number and the local \emph{tetraquark} number in the limit $m=h \gg J,K$ as a function of time, $Jt$, for the $3$-meson initial state in a chain of length $L=100$. We only show the central 50 sites. {\bf Top panel}: Data for $K=0.1<J^2/h$ where we qualitatively see long-lived repulsively bound \emph{tetraquark} and $3$-meson state. {\bf Middle panel}: Data for $K=0.3 \sim J^2/h$. In this regime, we see greater weight on the \emph{tetraquark} state and partial decay to the continuum at longer time. {\bf Bottom panel}: Data for $K=0.8>J^2/h$. In this regime, we see simple oscillations of bound $3$-meson and \emph{tetraquark} with negligible decay.}
    \label{fig:density_3m}
\end{figure}

We separate the Hamiltonian into two parts, the bare $H_0$ (the $h$ and $m$ terms) and the perturbative $H_1$ (the $J$ and $K$ terms). To obtain some simple insight into the physics, we first consider just the three states $\ket{m_3}$, $\ket{q_4}$ and $\ket{m_{1,r=2}}$ having bare energies $E_{m_3} = \bra{m_3}H_0\ket{m_3} = 2m+6h$, $E_{q_4} = \bra{q_4}H_0\ket{q_4} = 4m+4h$, and $E_{m_{1,r=2}} = \bra{m_{1,r=2}}H_0\ket{m_{1,r=2}} = 4m+4h$. At resonance where $h=m$, these states are degenerate under $H_0$. (The other states degenerate under $H_0$, $\ket{m_{1,r>2}}$ are necessary for a full treatment, and will be considered momentarily.)  The leading order coupling between the states $\ket{m_3}$ and $\ket{q_4}$ is given by $K$ (a single spin-flip) and between $\ket{q_4}$ and $\ket{m_{1,r=2}}$ is given by $-J^2/2h$ (a second-order meson hopping process \cite{supp}). In the subspace of these three states, the Hamiltonian is
\begin{equation}
\hat{\mathcal{H}} =
\begin{blockarray}{rccc}
 & \ket{m_3} & \ket{q_4}  & \ket{m_{1,r}} \\
\begin{block}{r[ccc]}
\bra{m_3}     & 2m+6h   & -K       & 0 \bigstrut[t] \\
\bra{q_4}     & -K       & 4m+4h   & -\tfrac{J^2}{2h} \\
\bra{m_{1,r}} & 0       & -\tfrac{J^2}{2h} & 4m+4h \bigstrut[b] \\
\end{block}
\end{blockarray}\,\,\,.
\label{eq:phn_hamil}
\end{equation}
Diagonalizing this Hamiltonian in the limit where $K \gg J^2/2h$ shows that the states $\ket{\pm} =  \ket{m_3}\pm\ket{q_4}$ are gapped out by energy $K$ from $\ket{m_{1,r=2}}$ and consequently, all of the continuum of separated $1$-meson states. These eigenstates correspond to bound \emph{tetraquark} states stabilized by a resonant particle number fluctuation, with the state at energy $-K$ being the conventional bound state and at energy $+K$ being the repulsively bound state. If we start from the $\ket{m_3}$ or the $\ket{q_4}$ state \cite{supp}, we expect coherent oscillations between these two states and no dissociation to the continuum (represented in this simplified model by the $\ket{m_{1,r}}$ state), showing an infinitely long lifetime of the \emph{tetraquark} state. Fig.~\ref{fig:density_3m} shows numerical results, discussed momentarily, that corroborate this basic picture.

To go beyond the $K \gg J^2/2h$ limit, we extend this picture to allow center-of-mass motion of the composite states, deriving the  effective model to second order in $J/h$ and $K/h$ shown in Fig.~\ref{fig:phenmlgy}(b). The center-of-mass position of all the states is labeled by $c$, and the relative separation between two $1$-mesons is labeled by $r$. Notations for the relevant states of interest are  \emph{tetraquarks}${}\equiv{}$$\ket{q_4^c}$, $3$-mesons${}\equiv{}$$\ket{m_3^c}$, and separated $1$-mesons${}\equiv{}$$\ket{m_1^{c,r}}$. A \emph{tetraquark} can dissociate to two $1$-mesons. Starting in a \emph{tetraquark}, one of the two mesons can then hop a distance $r$ and increase the separation by a distance $r$. If the other meson subsequently follows, a \emph{tetraquark} can re-form at a new center-of-mass position $c = r$. It can then also resonantly couple to a $3$-meson at that new position. We neglect center of mass tunnelings of $3$-mesons which are $\order{J^6/h^5}$~\cite{Meson_dynamics_Vaibhav}. Figure~\ref{fig:phenmlgy}(b) shows the off-diagonal matrix elements connecting these states, and all the second order processes in our effective model. Black arrows denote the matrix element, $K$ coupling a $3$-meson and a \emph{tetraquark} at the same position. Red arrows show the second-order degenerate perturbation theory matrix element, $-J^2/2h$. This encodes the transition of a \emph{tetraquark} to two separated $1$-mesons, and hopping of a $1$-meson that changes the separation between two $1$-mesons from $r$ to $(r \pm 1)$. Due to hopping of $1$-mesons, the $1$-meson states with different separations $r$ form a continuum of bandwidth $4J^2/h$.

Additionally, there are diagonal matrix elements that lift the degeneracies among these states. The \emph{tetraquarks} acquire a diagonal term, $-J^2/h+K^2/2h$, from both hopping and particle-number fluctuations. Similarly, the fluctuations in the separated $1$-mesons lead to a larger diagonal term $-2J^2/h+3K^2/4h$. These are derived in the Supplemental Material \cite{supp}. Defining $\mathcal{J}\equiv J^2/2h$ and $\mathcal{K}\equiv K^2/4h$, the effective model Hamiltonian is
\begin{align}
    \hat{\mathcal{H}} =  &\sum_c \Bigg[E_q \ket{q^c_{4}}\bra{q^c_{4}} + E_m  \sum_{r} \ket{m^{c,r}_{1}}\bra{m^{c,r}_{1}} \nonumber\\ &+ \Big\{K \ket{m^c_{3}}\bra{q^c_{4}}
    - \mathcal{J} \big( \ket{q^c_{4}}+\ket{q^{c+1}_{4}}\big)\bra{m^{c,2}_{1}}\nonumber\\ &-\mathcal{J}\sum_{r} \big( \ket{m^{c,r}_{1}}+\ket{m^{c+1,r}_{1}}\big)\bra{m^{c,r+1}_{1}}+\text{H.c.}\Big\}\Bigg], 
    \label{eq:eff_model}
\end{align}
where $E_q=2(-\mathcal{J}+\mathcal{K})$ and $E_m = (-4\mathcal{J}+3\mathcal{K})$.

When $K \ll J^2/h$, the $3$-meson and \emph{tetraquark} states are at a higher energy compared to the $1$-meson states in the continuum as shown in Fig.~\ref{fig:phenmlgy}(b). A finite value of $K$ creates a repulsively bound state that is a superposition of $3$-meson and \emph{tetraquark} states. The value of $K$ sets the amplitude of \emph{tetraquark} in this superposition. This state lies above the $1$-meson continuum of bandwidth $4J^2/h$, inhibiting its decay. We will quantitatively show in our results that this leads to a finite long-time probability of observing the \emph{tetraquark} state. In the Supplemental Material \cite{supp}, we show the energy spectrum of the spin model in Eq.~\eqref{eq:H_spin} and the effective model in Eq.~\eqref{eq:eff_model}, corroborating our claims in both the small- and large-$K$ regimes. We also show the \textit{tetraquark} initial state remains bound for all $K$.

\textbf{\textit{Quench dynamics.---}}We now present numerical simulation results supporting the phenomenological picture. 
The spin chain is initialized in a simple product state consisting of a $3$-meson at the center of the chain. The initial state is time-evolved under the Hamiltonian in Eq.~\eqref{eq:H_spin} using time-evolving block decimation (TEBD). We employ a Trotterized circuit with time step $J\delta t = 0.025$ and maximum bond dimension $\chi=32$; convergence with respect to these parameters is shown in the Supplemental Material \cite{supp}. This initial state is both experimentally accessible and ideally suited to capture the physics of interest using dynamics\cite{Meson_dynamics_Vaibhav}. We compute the local particle number, local \emph{tetraquark} number, and local $3$-meson number, defined as $\langle \hat{n}_i \rangle = \frac{1}{2} ( 1-\hat{\tau}_{i-1}^z\hat{\tau}_i^z)$, $\langle \hat{q}_{4i}\rangle = \langle \hat{n}_i \hat{n}_{i+1}\hat{n}_{i+2}\hat{n}_{i+3}\rangle$, and $\langle \hat{m}_{3i} \rangle = \langle \hat{n}_i (1-\hat{n}_{i+1}) (1-\hat{n}_{i+2}) \hat{n}_{i+3} \rangle$ respectively. Here, $i$ is the lattice site index, not to be confused with $c$ in the effective model. We also compare our TEBD results to exact diagonalization of the effective model in Eq.~\ref{eq:eff_model} where we fix $h = 10J$.

For our $L=100$ site chain, the initial state has one domain wall (particle) between spin sites 48-49 and another between spin sites 51-52 -- this is the $3$-meson initial state at the center. A tetraquark state at the center will have two additional domain walls between sites 49-50 and sites 50-51. To isolate the relevant physics, we also evolve the naive vacuum state (all spins down and no domain walls) and subtract its contribution from the data. This removes background excitations generated by the $K$-term, which can produce particle pairs in empty regions of the chain but are unrelated to the bound-state dynamics \cite{supp}.

\begin{figure}
    \centering
    \includegraphics[width=7.2cm]{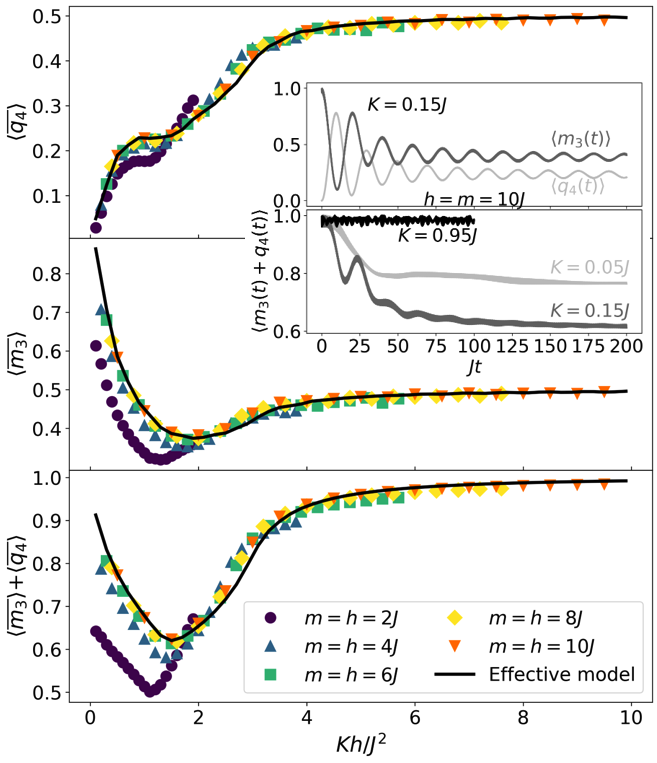}
    \caption{ Long-time average of the total \emph{tetraquark} number $\langle \overline{\hat{q}_{4}}\rangle$ and the total $3$-meson number $\langle \overline{\hat{m}_{3i}}\rangle$ as a function of $Kh/J^2$. The data points are computed using TEBD simulation of the full Hamiltonian while the black curve denotes simulation of the effective model. We see that as we increase $m=h$, the TEBD data approaches the effective model. \textbf{c.} The plot of $\langle \overline{\hat{m}_{3i}}\rangle+\langle \overline{\hat{q}_{4i}}\rangle$ shows that as we increase $Kh/J^2$, we nonmonotonically go from a bound hadronic state (repulsively bound) to enhanced dissociation to bound hadronic state again. {\bf Inset}: Time series plots of $\langle \hat{q}_4\rangle$ and $\langle \hat{m}_3\rangle$ for different values of $K$ and setting $h=m=10J$, showing a long-time asymptote along with small oscillations. For $m=h=8J,10J$ and $K\leq 0.25J$, TEBD calculations are carried  out to $T=200J^{-1}$ because of slow oscillations.}
    \label{fig:long_t}
\end{figure}

Figure \ref{fig:density_3m} shows particle and \emph{tetraquark} number density dynamics across the system for the $3$-meson initial state. For $h=m=6J \gg J,K$ with $K\gtrsim 8J^2/2h$ (bottom panel), we observe the dynamics to be almost entirely simple oscillations between the $3$-meson and \emph{tetraquark} states. As argued earlier both of these states are separated by an energy gap controlled by $K$ from the continuum. Reducing $K$ allows $J^2/2h$-driven hopping, producing a light-cone–like spreading that signals a fraction of the initial state dissociating into two separated $1$-mesons, while some finite fraction continues to oscillate between $\ket{m_3}$ and $\ket{q_4}$. Further decreasing $K$ (top panel) somewhat suppresses dissociation, yielding a fainter light cone, and a significant fraction of the initial state remains in a repulsive bound state. In what follows, we quantitatively study the long-time behavior of this bound state and dissociation as reflected in the observables of interest.

\textbf{\textit{Long-time behavior.---}}In this section, we study the long-time behavior of the model to understand the physics in different regimes. At long times, the total \emph{tetraquark} ($\langle \hat{q}_4\rangle = \sum_i \langle \hat{q_{4i}}\rangle$) and the total $3$-meson number ($\langle \hat{m}_3\rangle = \sum_i\langle \hat{m_{3i}} \rangle$) show oscillations about a finite value (\emph{inset} of Fig.~\ref{fig:long_t}). We compute the long-time averages of the total \emph{tetraquark} number ($\langle \overline{\hat{q}_4}\rangle$) and total $3$-meson number ($\langle \overline{\hat{m}_3}\rangle$) using an averaging scheme described in \cite{supp}.

The top and middle panels in Figure~\ref{fig:long_t} shows $\langle \overline{\hat{q}_4}\rangle$ and $\langle \overline{\hat{m}_3}\rangle$ as a function of $Kh/J^2$ from both TEBD simulations and the effective model simulations. We vary $K$ from $0.05J$ to $0.95J$. The effective model matches well with the TEBD data for large $m=h$. For large $K$,  we see the expected oscillations between the $\ket{q_4}$ and the $\ket{m_3}$ states with 50\% occupation of each of the states. The collapse of the TEBD data as a function of $K/(J^2/h)$ at large $m=h$ shows that $K$ and $J^2/h$ are the relevant energy scales, and the crossover between large-$K$ and small-$K$ regions is set by $K\sim J^2/h$.

The bottom panel of Fig.~\ref{fig:long_t} plots $\langle \overline{\hat{q}_4}\rangle+\langle \overline{\hat{m}_3}\rangle$, showing that for small $Kh/J^2$, there is a bound state with increasing \emph{tetraquark} composition as $K$ increases. In this regime, there is minimal dissociation to the continuum since the bound state is pushed above the band due to contributions from quantum fluctuations. The effective model structure in Fig.~\ref{fig:phenmlgy}(b) captures the repulsive bound state physics arising due to the finite bandwidth of $1$-mesons induced by the discrete nature of the lattice gauge theory. This physical phenomena is absent in a continuum gauge theory.

The bound state probability changes  nonmonotonically  with $K$. Increasing $K$, we go from minimal dissociation (repulsive bound states) to notable dissociation to again no dissociation (due to a gap of order $K$). This causes the long-time $\langle \overline{\hat{q}_4}\rangle+\langle \overline{\hat{m}_3}\rangle$ number to be large for small $Kh/J^2$ followed by a dip and finally increasing to $1$. The dissociation is maximized in the intermediate regime when $K \sim J^2/h$, and is enhanced further as $h=m$ decreases.

In the regime where $h=m \sim J$, the effective model is insufficient as fluctuations can create longer mesons not considered in the effective model. Presence of these excitations lowers the values of $\langle \overline{\hat{q}_4}\rangle$ and $\langle \overline{\hat{m}_3}\rangle$. This is captured by the TEBD data in Fig.~\ref{fig:long_t} as we go to smaller values of $m=h$. 

\textbf{\textit{Summary and outlook.---}}We have observed the dynamical production of repulsive bound states of two mesons by time evolving a simple high-energy product state according to the particle non-conserving $\mathbb{Z}_2$ lattice gauge theory Hamiltonian, Eq.~\eqref{eq:H_spin}. These bound states appear solely due to quantum fluctuations and particle pair-production processes that inhibit their decay into a low-energy continuum of states. We quantitatively analyze this process by performing numerical simulations using tensor network methods. We further derive an effective model whose predictions agree with the numerical simulations.

In contrast to earlier works~\cite{hadron1,hadron2}, we have observed that two mesons can bind together without requiring any finite-range matter interaction terms. The quantum fluctuations of the gauge fields are crucial contributors to the energy of this hadron state, reminiscent of how gluon fluctuations significantly contribute to the mass of protons and neutrons in the theory of strong nuclear force. The phenomenology we show here can be generalized to more complex hadronic bound states, such as ones involving more than two $1$-mesons at higher energies in the spectrum. It would be interesting to explore whether these binding mechanisms are stable with gauge-invariant higher-order perturbations, theories with different gauge symmetry groups, and higher spatial dimensions. It would also be interesting to study the nonequilibrium dynamics resulting from scattering these high-energy hadrons to study analogs of high-energy collisions. Finally, an intriguing possibility to explore is whether metastable repulsively bound states can occur in materials with emergent gauge degrees of freedom.

\bigskip
\footnotesize
\begin{acknowledgments}
S.G.R. would like to thank Kevin Slagle and Fang Xie for invaluable insights. V.S.~acknowledges support from the J.~Evans Attwell-Welch fellowship by the Rice Smalley-Curl Institute.
K.R.A.H., S.G.R, and V.S.~acknowledge support from the W.~M.~Keck Foundation (Grant No.~995764), the Welch Foundation (C-2166), the Office of Naval Research
(N00014-20-1-2695), and the Department of Energy (DE-SC0024301). K.X., U.B., and J.C.H.~acknowledge funding by the Max Planck Society, the Deutsche Forschungsgemeinschaft (DFG, German Research Foundation) under Germany’s Excellence Strategy – EXC-2111 – 390814868, and the European Research Council (ERC) under the European Union’s Horizon Europe research and innovation program (Grant Agreement No.~101165667)—ERC Starting Grant QuSiGauge. Views and opinions expressed are, however, those of the author(s) only and do not necessarily reflect those of the European Union or the European Research Council Executive Agency. Neither the European Union nor the granting authority can be held responsible for them. This work is part of the Quantum Computing for High-Energy Physics (QC4HEP) working group. This work was supported in part by the Big-Data Private-Cloud Research Cyberinfrastructure MRI-award funded by NSF under grant CNS-1338099 and by Rice University's Center for Research Computing (CRC).
\end{acknowledgments}
\normalsize

\bibliography{main_text}

\end{document}


\preprint{APS/123-QED}

\title{Supplemental Material: \\ Repulsively Bound Hadrons in a $\mathbb{Z}_2$ Lattice Gauge Theory}

\author{Sayak Guha Roy${}^{\orcidlink{0000-0001-5816-6079}}$}
\email{sg161@rice.edu}
\affiliation{Department of Physics and Astronomy, Rice University, Houston, TX 77005 and Smalley-Curl Institute, Rice University, Houston, TX 77005}

\author{Vaibhav Sharma${}^{\orcidlink{0000-0003-3151-2968}}$}
\affiliation{Department of Physics and Astronomy, Rice University, Houston, TX 77005 and Smalley-Curl Institute, Rice University, Houston, TX 77005}

\author{Kaidi Xu${}^{\orcidlink{0000-0003-2184-0829}}$}
\affiliation{Max Planck Institute of Quantum Optics, 85748 Garching, Germany}
\affiliation{Munich Center for Quantum Science and Technology (MCQST), 80799 Munich, Germany}

\author{Umberto Borla${}^{\orcidlink{0000-0002-4224-5335}}$}
\affiliation{Max Planck Institute of Quantum Optics, 85748 Garching, Germany}
\affiliation{Munich Center for Quantum Science and Technology (MCQST), 80799 Munich, Germany}

\author{Jad C.~Halimeh${}^{\orcidlink{0000-0002-0659-7990}}$}
\affiliation{Department of Physics and Arnold Sommerfeld Center for Theoretical Physics (ASC), Ludwig Maximilian University of Munich, 80333 Munich, Germany}
\affiliation{Max Planck Institute of Quantum Optics, 85748 Garching, Germany}
\affiliation{Munich Center for Quantum Science and Technology (MCQST), 80799 Munich, Germany}
\affiliation{Department of Physics, College of Science, Kyung Hee University, Seoul 02447, Republic of Korea}

\author{Kaden R.~A.~Hazzard${}^{\orcidlink{0000-0003-2894-7274}}$}
\affiliation{Department of Physics and Astronomy, Rice University, Houston, TX 77005 and Smalley-Curl Institute, Rice University, Houston, TX 77005}

\maketitle

\section{Numerical Simulation}
We use the time-evolving block decimation (TEBD) algorithm applied to  matrix product states \cite{Vidal1,MPS_brickwork} to study the time evolution of the $\mathbb{Z}_2$ lattice gauge theory spin model of Eq. 2 of the main text. MPS-based time evolution techniques have been significantly used in literature \cite{MPS_brickwork,ORUS2014117,Orus2019}. We initialize the spin chain in one of two different initial states, (i) a $3$-meson state at the center of the chain, and (ii) a \emph{tetraquark} state at the center of the chain. These are simple product states, shown in Fig. 1 of the main text, that can be represented by a bond dimension $\chi=1$ MPS. For the TEBD time evolution, we employ a Trotterized circuit with Trotter step $J\delta t=0.025$ and maximum bond dimension $\chi=32$. We show here that this leads to well-converged results.
 
\begin{figure}
    \centering
    \includegraphics[width=15cm]{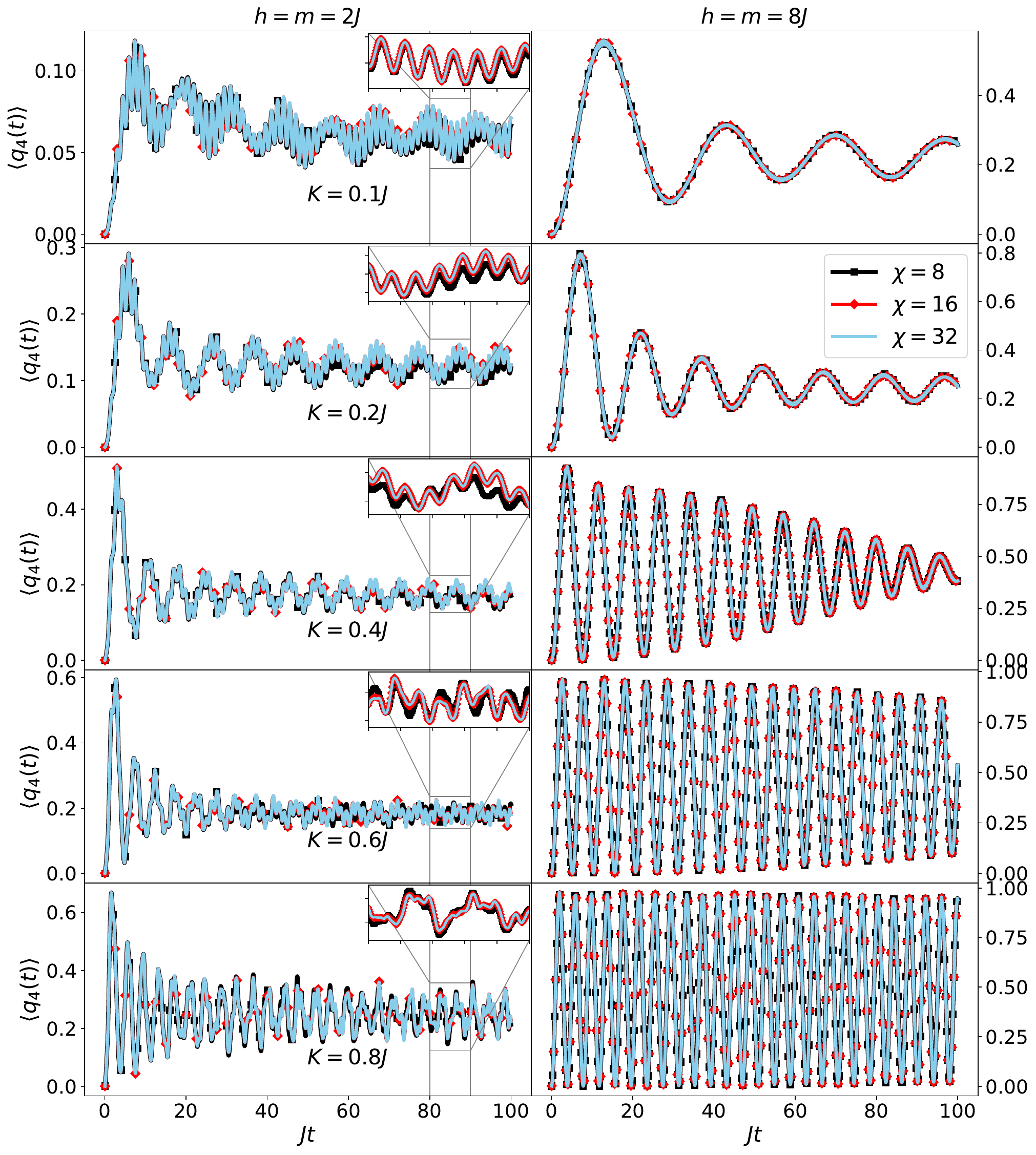}
    \caption{Total \emph{tetraquark} number as a function of time, $\langle q_4(t)\rangle$, for $m=h=2J$ (left panel) and $m=h=8J$ (right panel) using TEBD simulations with different bond dimension cutoffs, $\chi$ for different model parameter values. We see that the curves are well-converged for $\chi=32$, the value used for all calculations in the main text.}
    \label{fig:chi_error}
\end{figure}

For the $\mathbb{Z}_2$ lattice gauge theory Hamiltonian and the initial states that we have considered, a small value of $\chi$ accurately captures the dynamics because of the relatively low entanglement growth in our dynamics, as  seen in  Fig.~\ref{fig:chi_error}, which plots the total \emph{tetraquark} number as a function of time for different model parameter values and different values of $\chi$. These clearly show that $\chi=32$ gives sufficient accuracy to capture the dynamics being studied. This is expected, since the states appearing are largely composed of a relatively small number of particles. We have checked that the total $3$-meson number converges similarly. 

\begin{figure}
    \centering
    \includegraphics[width=15cm]{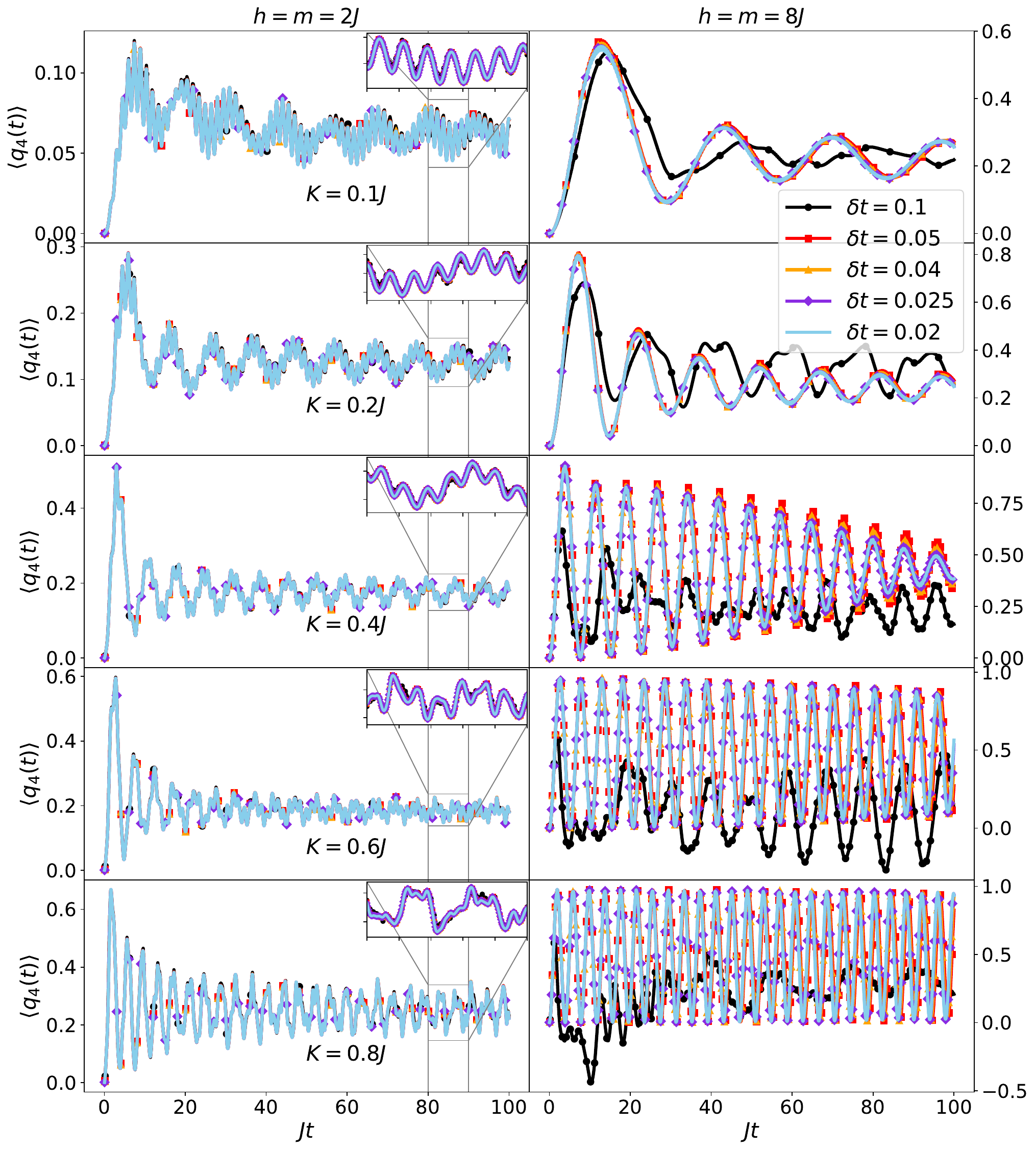}
    \caption{Total \emph{tetraquark} number as a function of time, $\langle q_4(t)\rangle$, for $m=h=2J$ (left panel) and $m=h=8J$ (right panel) using TEBD simulations with different Trotter time steps $J\delta t$. A $J\delta t=0.025$  is accurately converged and is used for all calculations in the main text.}
    \label{fig:dt_error}
\end{figure}

Additionally, Trotterizing the time dynamics leads to Trotter error \cite{PhysRevX.11.011020}, which decreases with increasing number of Trotter steps, each corresponding to a time step $\delta t$. In Fig.~\ref{fig:dt_error}, we plot the total \emph{tetraquark} number as a function of time for different model parameter values and different values of $\delta t$. We see that the curves are well converged for $J\delta t \leq 0.025$. The total $3$-meson number shows similar convergence behavior. We therefore choose $\delta t=0.025$ for our calculations  in the main text.

\subsection{Background Subtraction}

The true vacuum of the particle non-conserving $\mathbb{Z}_2$ lattice gauge theory differs from the simple no-matter state that is used as the background for our initial state, onto which the \emph{tetraquark} or 3-meson states are added. 
In the no-matter vacuum state, the pair-production term ($K$) causes particle-number fluctuations. We call these fluctuations background effects as they show up uniformly throughout the system including regions far from any matter. They are most prominent when $K/J$ is large and $m/J$ and $h/J$ are small, While we find that these background fluctuations have little effect on the bound state physics in the parameter regimes we consider, it does cause small but noticeable  fast oscillations that can be distracting for analysis.

\begin{figure}
    \centering
    \includegraphics[width=12cm]{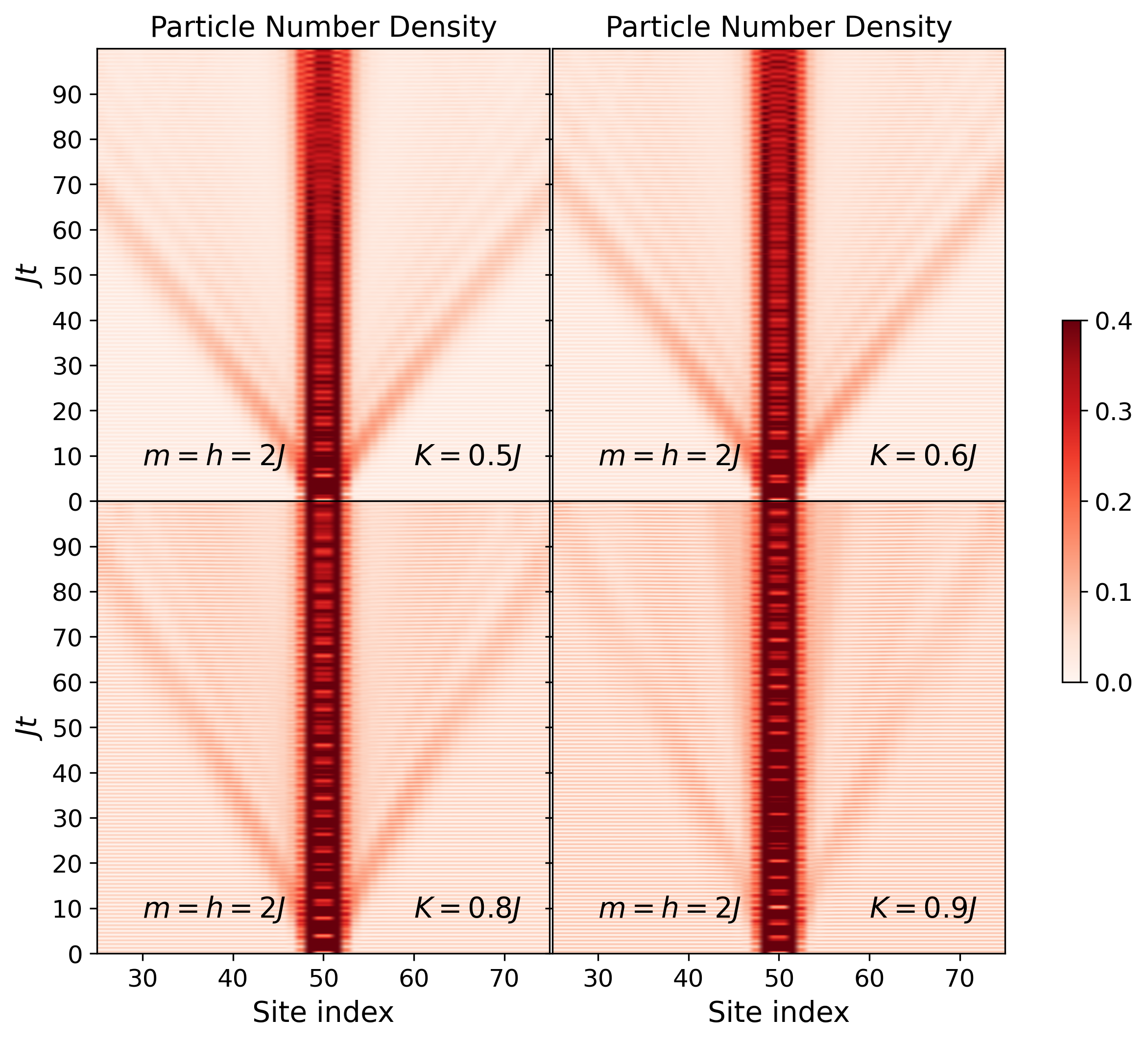}
    \caption{Particle number density as a function of site index and time in a spin chain of length $L=100$, showing sites $25$ to $75$ for the $3$-meson ($\ket{m_3}$) initial state without vacuum background subtraction. The background effects are seen as the horizontal stripes that get more prominent as $K$ increases.}
    \label{fig:no_vac}
\end{figure}

To remove the background dynamics, we additionally time-evolve the particle vacuum state (all spins down initial state) and subtract its results from those of the actual simulation. In Fig.~\ref{fig:no_vac}, we show number density plots for varying values of $K/J$ without background subtraction. We can
see uniform horizontal stripes that become more prominent as $K/J$ increases. These are the background fluctuations. Our background subtraction removes these stripes as we can see from the number density plots in Fig. 2 in the main text. Note that here in Fig.~\ref{fig:no_vac}, we are only showing the density data between sites $25$ and $75$, hence the light cones which seem to hit the boundary are not actually hitting the boundaries.

\subsection{Time averaging}

In Fig. 3 of the main text, we present long-time averages of the total tetraquark and three-meson numbers. The corresponding timeseries data, shown in the inset of Fig. 3 of the main text and in Figs. \ref{fig:chi_error} and \ref{fig:dt_error}, display persistent oscillations about a well-defined mean. In this section, we show that despite the persistent oscillations, the dynamics support a convergent long-time average. After an initial transient period, the system enters a long-lived oscillatory regime in which the cumulative average stabilizes when the averaging window is chosen within this late-time behavior, yielding the values reported in the main text.

\begin{figure}
    \centering
    \includegraphics[width=0.75\linewidth]{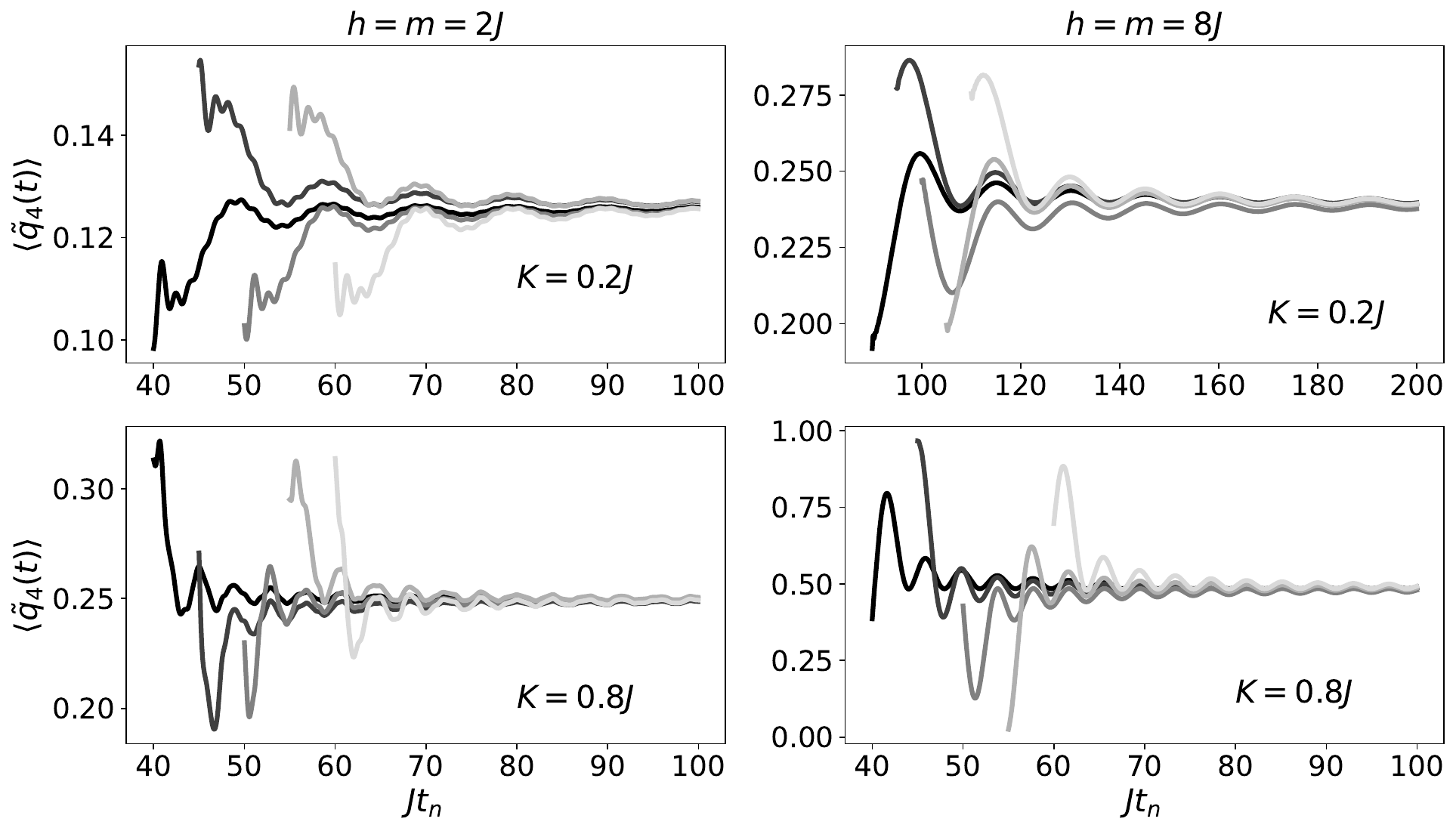}
    
    \caption{Cumulative average of the total \emph{tetraquark} number time series data shown in Eq. \ref{eq:c_av} as a function of $t_n$ for different parameter regimes. The different shaded curves are for different start times of the averaging ($T_0=t_{k_0}$). $T_0$ for different curves can be interpreted from where the curves start. We see that the long time average of the total \emph{tetraquark} number asymptotes to a constant value which converges for different choices of start times.}
    \label{fig:tqe_av_conv_1}
\end{figure}

The long-time average of a time-dependent observable $f(t)$ is  
\begin{equation}
    \overline{\langle f\rangle} = \lim_{T\rightarrow \infty}\frac{1}{T-T_0} \int_{T_0}^T \langle f(t)\rangle \text{d}t\,.
    \label{eq:av}
\end{equation}
Here $T$ is the total time of our simulations and $T_0$ is a cutoff chosen at sufficiently long time such that the long time average  converges. To explicitly assess convergence, we consider the cumulative average of a finite time series,
\begin{equation}
    \langle \Tilde{f}(t_n)\rangle = \frac{1}{n} \sum_{k=k_0}^{n} \langle f(t_k) \rangle,
    \label{eq:c_av}
\end{equation}
were, $t_n$ is the time at index $n^{\text{th}}$ step; $k_0$ labels the starting of the averaging window, and $T_0 = t_{k_0}$. 

Figure \ref{fig:tqe_av_conv_1} shows the cumulative average of the total \emph{tetraquark} number as a function of $t_n$ for several choices of the averaging start time $T_0 = t_{k_0}$ and various parameter values. The different curves correspond to values of $T_0$ chosen within the later portion of the time series, after the initial transient dynamics have subsided. For each choice of $T_0$, the cumulative average approaches a stable value at long times, with the final value corresponding to the long-time average defined in Eq. \eqref{eq:av}. While the extracted average depends on the choice of $T_0$ if the averaging window includes early-time dynamics, we find that for start times chosen sufficiently late in the evolution, the cumulative averages converge to the same asymptotic value. This behavior confirms that the long-time averages reported in the main text are representative of the persistent oscillatory regime rather than early-time dynamics. The specific choices of $T_0$ used for different parameter regimes are discussed below.

For small $K$ and large $h=m$, longer evolution times $Jt$ are required for the cumulative average to reach its asymptotic value. Accordingly, for $h=m=8J$ and $10J$ and for $K \leq 0.25J$, we evolve the system up to a total time $T = 200J^{-1}$ in the TEBD simulations. For smaller values of $h=m$ ($2J$, $4J$, and $6J$), the oscillations occur on shorter timescales, and the cumulative average stabilizes before $t = 50J^{-1}$; in these cases we therefore choose $T_0 = 50J^{-1}$. For $h=m=8J$ and $10J$, the choice of averaging start time depends on $K$. When $K \leq 0.25J$, we take $T = 200J^{-1}$ and average over the second half of the time series, corresponding to $T_0 = 100J^{-1}$. For larger values of $K$, we reduce the total evolution time to $T = 100J^{-1}$ and choose $T_0 = 50J^{-1}$ for $K = 0.3J$, $T_0 = 20J^{-1}$ for $K = 0.35J$, and $T_0 = 0J^{-1}$ for $K > 0.35J$, reflecting the faster approach to the persistent oscillatory regime. The effective model is simulated at $h=10J$, and we adopt the same values of $T_0$ as used for the corresponding $h=10J$ TEBD data with a total simulation time, $T = 200J^{-1}$ for all parameters.

\section{Effective model: Derivation of Matrix Elements}

In this section, we derive, by second-order degenerate perturbation theory in $J$ and $K$, the effective model Eq. 4, illustrated in Fig. 1(b) of the main text. The (zero'th order) degenerate manifold that is connected to our initial states are $\ket{m^c_{3}}$: all the $3$-meson states possible in the spin chain, $\ket{q^c_{4}}$: all the \emph{tetraquark} states possible in the chain, and $\ket{m^{c,r}_{1}}$: all the pairs of separated 1-meson states that are possible. Here, $c$ is the center of mass position and $r$ is the relative separation between two $1$-mesons. All these states are shown in Fig. 1(b) of main text for a length $L=6$ chain. 

\begin{figure}
    \centering
    \includegraphics[width=0.7\linewidth]{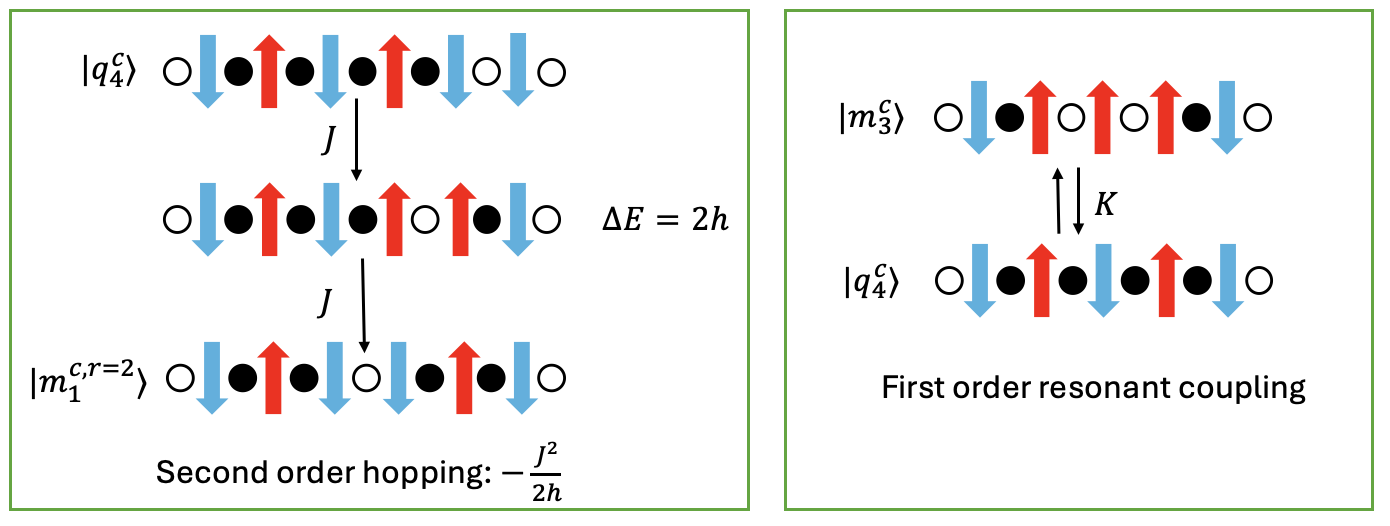}
    \caption{\textbf{Left panel}: Second-order hopping process coupling the $\ket{q^c_{4}}$ states with the nearest separated $1$-meson states $\ket{m^{c,r=2}_{1}}$. \textbf{Right panel}: Resonant first-order process coupling the $\ket{m^c_{3}}$ and the $\ket{q^c_{4}}$ states. }
    \label{fig:Eff_odiag}
\end{figure}

The only first-order term comes from the pair-production term coupling  $3$-meson states, $\ket{m^c_{3}}$,  to  \emph{tetraquark} states, $\ket{q^c_{4}}$, with matrix element $K$. This (with the rest of the off-diagonal perturbative processes) is shown in Fig.~\ref{fig:Eff_odiag}. 

The remaining terms are second order. The tetraquark states, $\ket{q^c_{4}}$, couple to the separated $1$-meson state, $\ket{m^{c,r=2}_{1}}$ by a second-order hopping process with matrix element $-J^2/2h$. Similarly, the $1$-meson states with relative separation $r$ and $r+1$ are coupled by the same second-order hopping process with the matrix element, $-J^2/2h$.

In addition to the off-diagonal elements, there are  $O(J^2/h)$     (hopping fluctuation) and $O(K^2/h)$ (particle creation and annihilation fluctuation) processes that lead to diagonal matrix elements. This raises the $3$-meson and \emph{tetraquark} states'   energy relative to the continuum of separated $1$-meson states -- as seen in the main text.

\begin{figure}
    \centering
    \includegraphics[width=0.9\linewidth]{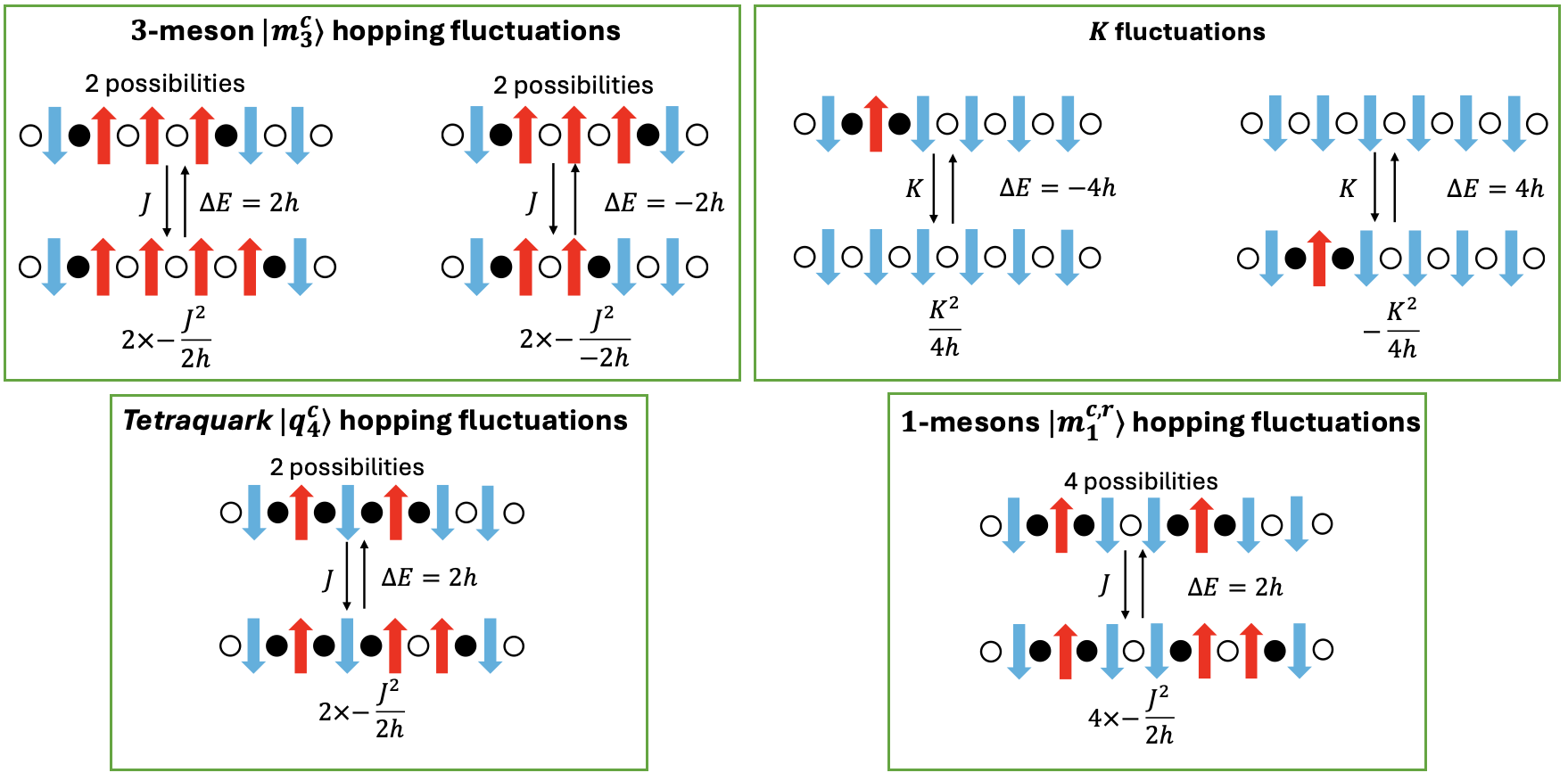}
    \caption{Hopping fluctuations, due to the  $J$ term, of the $3$-meson ($\ket{m_3^c}$), \emph{tetraquark} ($\ket{q_4^c}$), and the $1$-meson states ($\ket{m_1^{c,r}}$) along with their contribution to the diagonal matrix element. $K$ fluctuations are the particle creation and annihilation fluctuations, due to the $K$ term.}
    \label{fig:eff_diag}
\end{figure}

First we consider hopping-induced second-order processes. Each $3$-meson state couples to two different types of intermediate states -- $4$-mesons and $2$-mesons -- which have opposite energies relative to the $3$-meson, and therefore the processes cancel each other out, as shown in Fig.~\ref{fig:eff_diag}. The \emph{tetraquark} states on the other hand have processes only involving higher energy states, so their energy is shifted  by $-J^2/h$ relative to the $3$-meson. The separated $1$-meson states have four processes coupling to higher energy states, which lowers their energy by $2J^2/h$ relative to the $3$-meson state. 

Next, we consider second-order particle number fluctuation  processes, illustrated in Figure~\ref{fig:eff_diag}. The \emph{tetraquark} and $1$-meson states can first annihilate a pair and  then create it back, giving a $K^2/4h$ contribution to the diagonal matrix elements. All the states can create and then destroy a pair anywhere in the entire chain where there are empty neighboring sites. 
The $3$-meson and the \emph{tetraquark} states have $L-5$ such possible processes while the separated $1$-meson states have $L-6$ such processes in a chain of length $L$. Since the $1$-meson states have one less pathway, the relative energy of the $1$-meson continuum is shifted by $K^2/4h$.

\section{Tetraquark Initial State}

\begin{figure}
    \centering
    \includegraphics[width=10cm]{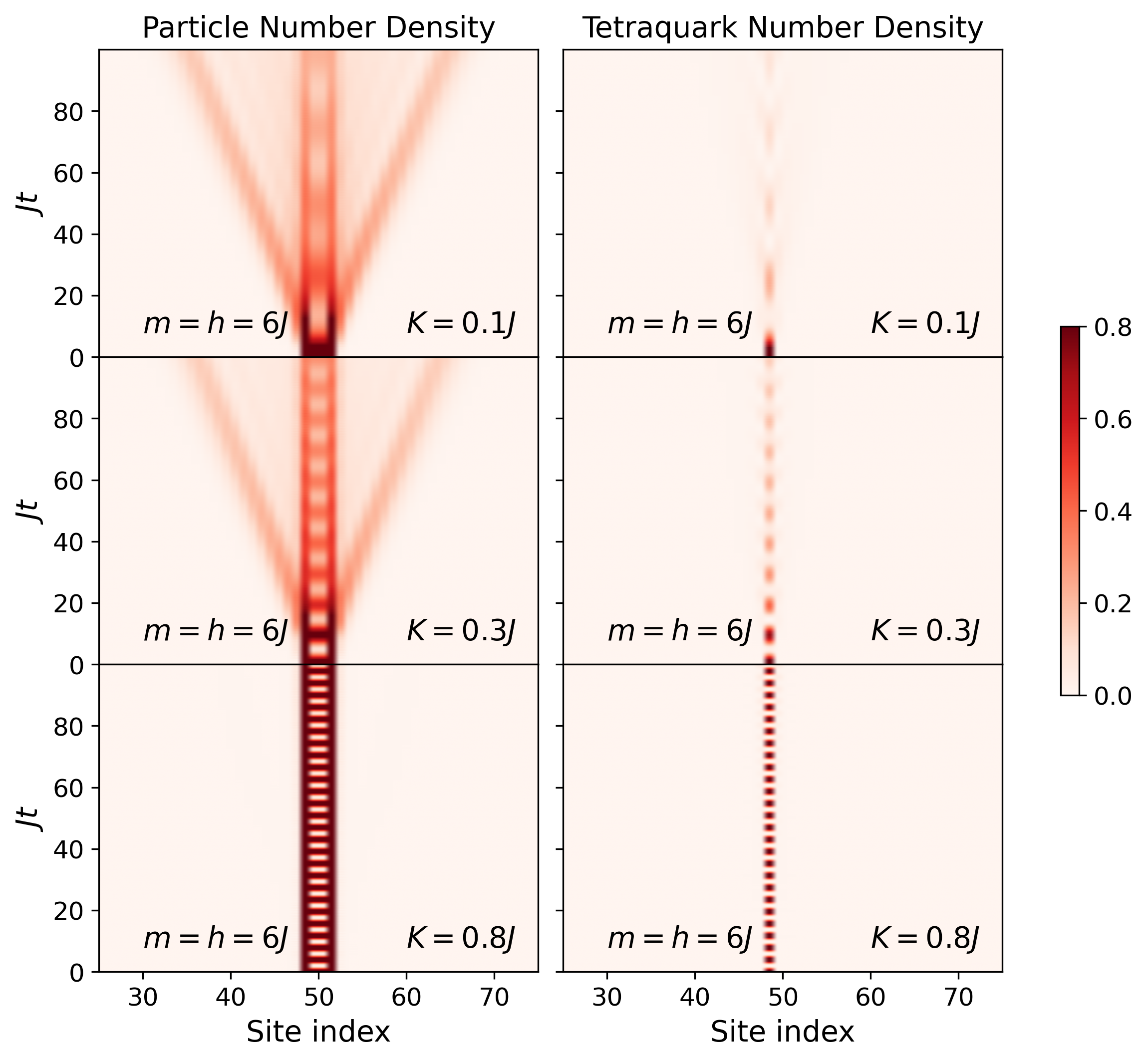}
    \caption{Particle density (left panel) and the total \emph{teraquark} number density (right panel) for the \emph{tetraquark} initial state in the center of the chain for differnt values of $K/J$ at $h=m=6J$.}
    \label{fig:density_tq}
\end{figure}
\begin{figure}
    \centering
    \includegraphics[width=14cm]{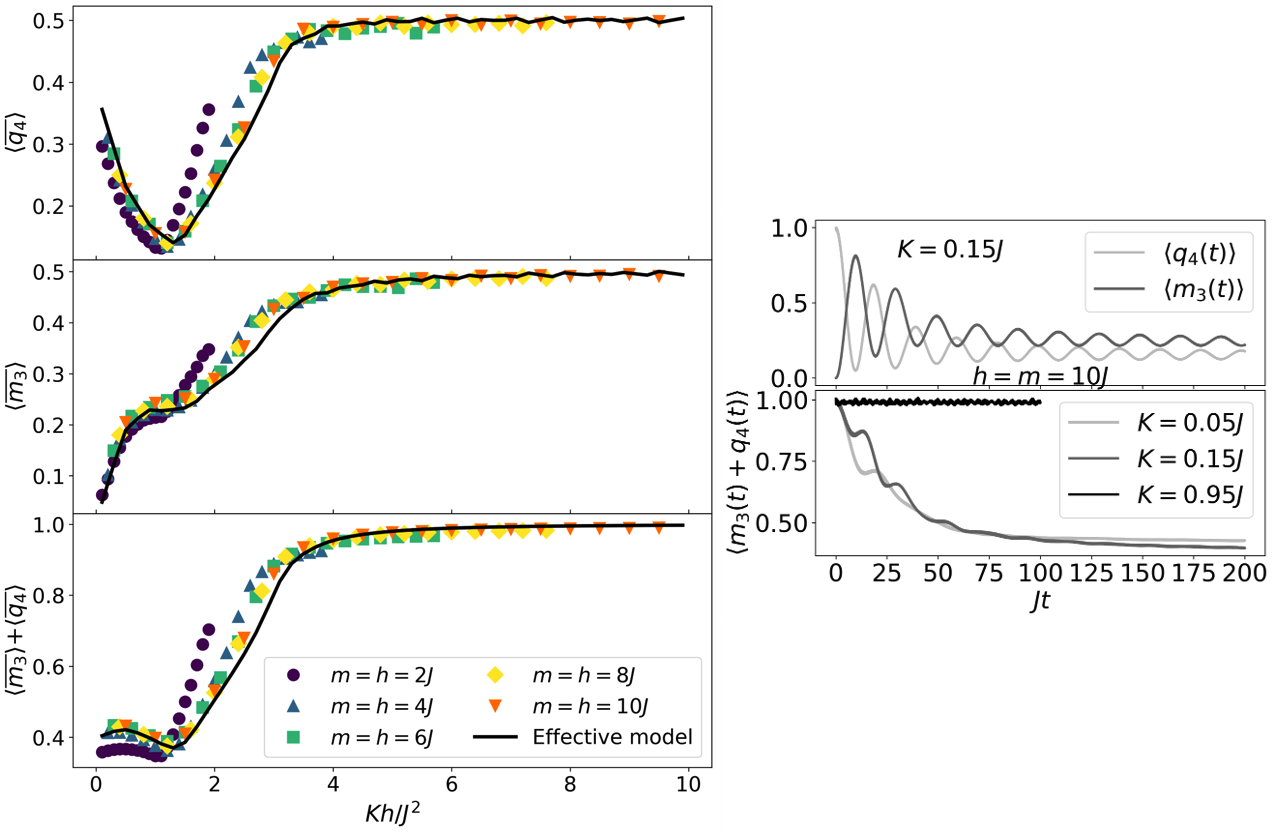}
    \caption{\textbf{Left}: Long-time average of the $3$-meson (top), \emph{tetraquark} (middle) and their sum (bottom panel) as a function of $Kh/J^2$ for the \emph{tetraquark} initial state. \textbf{Right}: Time series plots of the $3$-meson and \emph{tetraquark} number as a function of time for $h=m=10J$.}
    \label{fig:lt_tq}
\end{figure}

In the main text, we considered dynamics initiated from a $3$-meson state localized in the center of the $L=100$ spin chain. In this section, we will show that the dynamical signatures of our bound states also show up when we initialize our system with a \emph{tetraquark} at the center of the chain. Specifically, at large-$K$ the initial state leads to localized oscillations between the $3$-meson and \emph{tetraquark} state, reflecting the initial state's equal probability of occupying the attractive and repulsive bound states in this regime. For smaller $K$, some non-unity but finite fraction remains localized in a repulsively-bound state that is a superposition of a $3$-meson and a \emph{tetraquark} state at the origin. 

The large-$K$ regime is readily understood, since the gap between the states $ \ket{m_3}\pm\ket{q_4}$ and the $1$-meson continuum is set by $K$. When $K$ is larger than the half bandwidth of the continuum ($ 2J^2/h$), we observe both attractive and repulsive bound states at energy $\sim\pm K$ just like the $3$-meson initial state case. This is shown by the bottom panel of Fig.~\ref{fig:density_tq} where we observe oscillations between the $\ket{m_3}$ and the $\ket{q_4}$ states. 

In the low-$K$ regime, the repulsive bound state located above the continuum band has a greater overlap with the $3$-meson state compared to the \emph{tetraquark} state. Thus the initialized \emph{tetraquark} state has a  lower overlap with the repulsive bound state in contrast to the $3$-meson initial state. The energy difference between the \emph{tetraquark} states and the separated $1$-meson states is $J^2/h$ (Fig. 1(b) of the main text), which is less than the continuum's half bandwidth, $2J^2/h$. This causes the \emph{tetraquark} initial state to have a stronger hybridization with the continuum, leading to notable dissociation into the continuum compared to the $3$-meson initial state. The top and middle panels of Fig.~\ref{fig:density_tq} show the intermediate and the low-$K$ regime where we observe this dissociation. However, there is some finite long-time probability of observing the \emph{tetraquark} and $3$-meson localized near the center due to the initial state's finite overlap with the repulsive bound state.

Analogously to the $3$-meson case in Fig. 4 of the main text, we also compare long-time averages of the total \emph{tetraquark} ($\langle \overline{q_4}\rangle$) and $3$-meson number ($\langle \overline{m_3}\rangle$) with the effective model as shown in Fig.~\ref{fig:lt_tq}. We note that at long times for large $K$, the tetraquark number and the $3$-meson number asymptote to 0.5 signifying $50\%$ occupation of each states. In the plot of the sum of $\langle \overline{m_3}\rangle + \langle\overline{q_4}\rangle$, we do not see an upturn for small $K$ that we saw for the $3$-meson initial state. But we still have a non-zero long time \emph{tetraquark} and $3$-meson number, showing repulsive bound states along with some hybridization with the continuum. This hybridization is clearly greater for the \emph{tetraquark} initial state than the $3$-meson initial state. We still observe the dynamical signatures of our bound state even when we initialize a \emph{tetraquark} in the middle of the chain.

Note that Fig.~\ref{fig:lt_tq} shows that there is a finite long-time probability of observing the \emph{tetraquark} state even when $K \to 0$. In this limit, the \emph{tetraquark} initial state is completely decoupled from the $3$-meson state. The repulsive  bound state that is above the continuum is purely a $3$-meson state. Thus the \emph{tetraquark} initial state is only coupled to the continuum and as argued above, lies within the continuum bandwidth. Normally one would expect it to dissociate completely in this limit. However, in our case, the coupling ($J^2/2h$) to the continuum is comparable to the bandwidth ($4J^2/h$). This relatively large coupling causes some quasi-localization of the \emph{tetraquark} states, preventing its complete decay. 

One can understand this by interpolating between two limits. Let the coupling of the \emph{tetraquark} state to the continuum be denoted by $\kappa$ and the continuum bandwidth to be denoted by $\beta$. In the limit where $\kappa \ll \beta$, the \emph{tetraquark} state would completely decay in the long-time limit with a rate proportional to $\kappa^2$ consistent with the Fermi's golden rule. In the opposite limit where $\kappa \gg \beta$, the \emph{tetraquark} state would form a localized bound state with one of the continuum states of energy $\pm \kappa$. This would separate them from the rest of the continuum and show no decay. Our limit where $\kappa \sim \beta$ interpolates between these limits. The \emph{tetraquark} initial state shows some quasi-localization with a finite non-unity probability of survival at long times.

\section{Off resonance: $h \ne m$}

\begin{figure}
    \centering
    \includegraphics[width=0.5\linewidth]{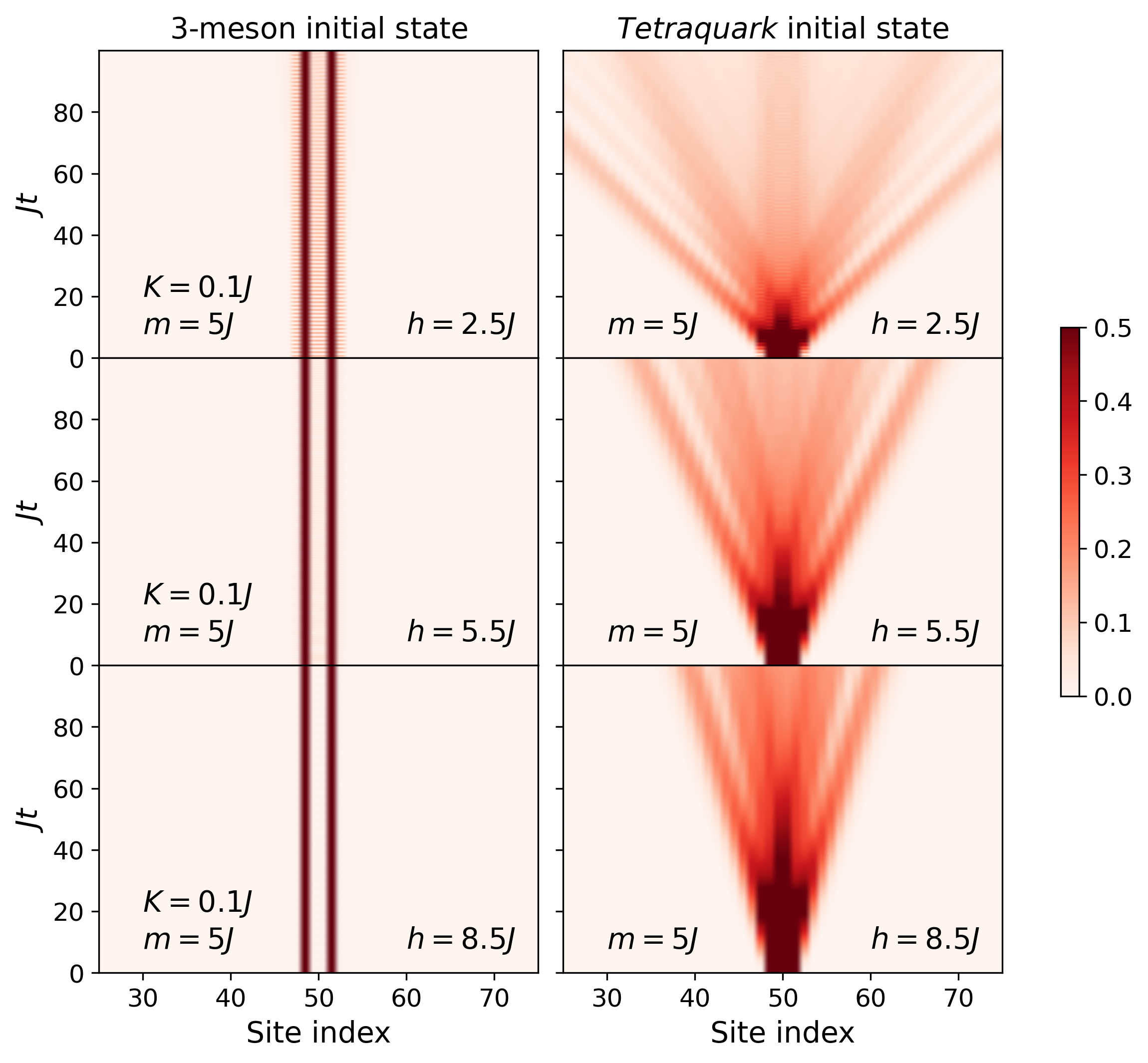}
    \caption{Total particle number density for the $3$-meson initial state (left) and the \emph{tetraquark} initial state (right) at off-resonance ($m \neq h$) for $m=5J$, $K=0.1J$, and various $h/J$. We see confinement of the $3$-meson initial state that delocalizes at the rate of $J^6/h^5$ \cite{Meson_dynamics_Vaibhav} and a notable dissociation of the \emph{tetraquark} initial state into $1$-mesons at the rate of $J^2/h$.}
    \label{fig:den_off_res1}
\end{figure}

We have shown phenomenologically that the repulsive bound state persists for all $K$ and that the regular (attractive) bound state physics exists at sufficiently large values of $K$ at resonance, i.e. when $h=m$. When $|h-m|$ is large we expect the picture to break down because the $3$-meson state and the \emph{tetraquark} states are no longer resonantly coupled. Thus there is no bound state that is separated from the continuum and contains a finite fraction of the \emph{tetraquark} state. Fig.~\ref{fig:den_off_res1} shows the dynamics of the total particle number density for both the $3$-meson and the \emph{tetraquark} initial states for small $K/J$ where repulsive bound states are expected to occur if $h=m$. The \emph{tetraquark} state never appears dynamically for the $3$-meson initial state and on the other hand, the $3$-meson state never appears for the \emph{tetraquark} initial state. The $3$-meson state remains localized for a long time due to the large $h/J$ term causing standard meson confinement that leads to slow delocalization of longer mesons~\cite{Meson_dynamics_Vaibhav}. The \emph{tetraquark} initial state quickly delocalizes to separated $1$-mesons at the rate of $J^2/h$. Therefore, at off resonance, we lose the  four-particle bound state physics caused by the resonant pair-production term. 

\begin{figure}
    \centering
    \includegraphics[width=0.5\linewidth]{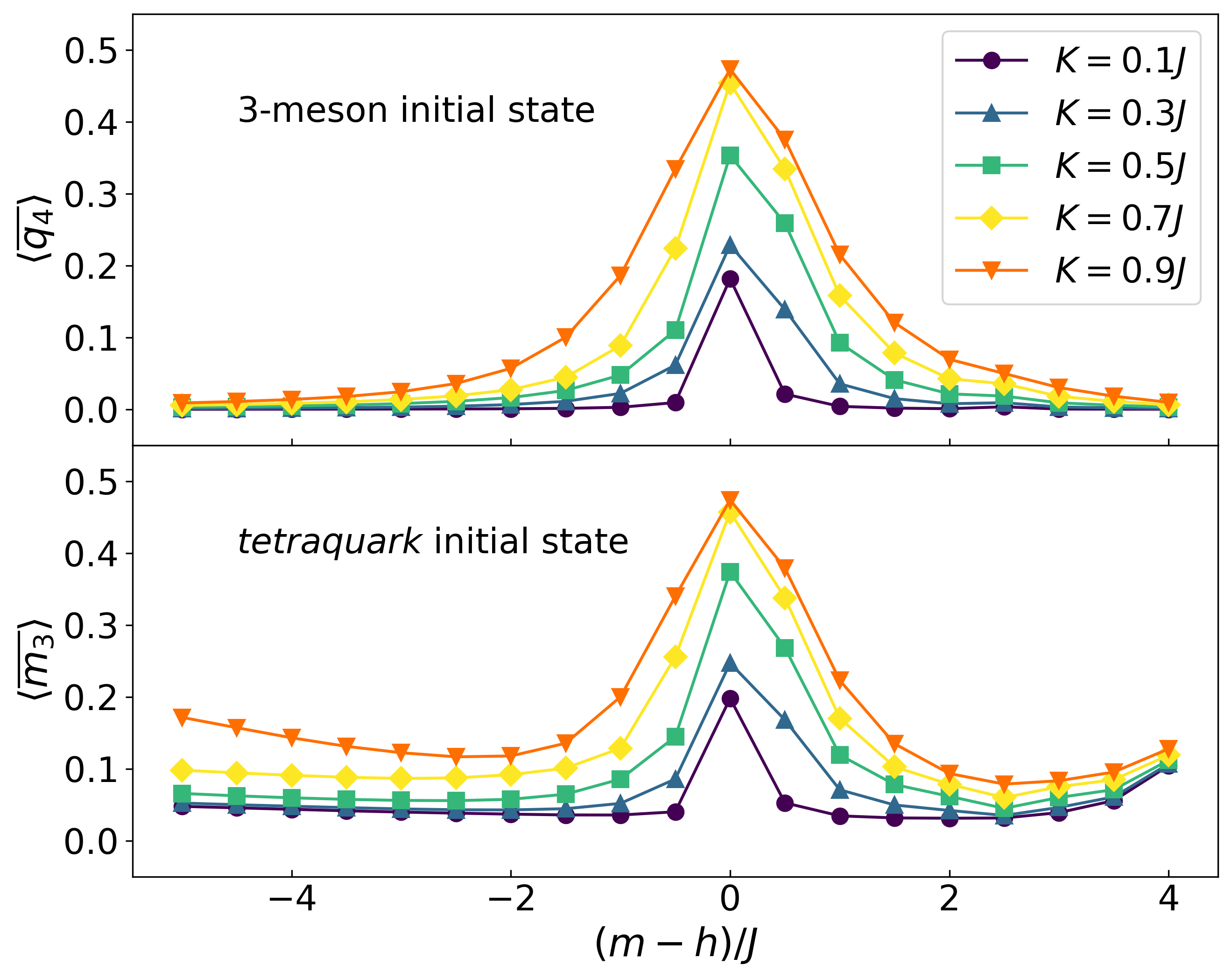}
    \caption{Long-time average of the total \emph{tetraquark} number for the $3$-meson initial state and the total $3$-meson number for the \emph{tetraquark} initial state as a function of $(m-h)/J$. We see the expected peak at $m-h=0$ at resonance and a sharp decay as we go away from resonance.}
    \label{fig:off_res}
\end{figure}

In Fig.~\ref{fig:off_res}, we plot the long-time average of the total \emph{tetraquark} number for the $3$-meson initial state and the total $3$-meson number for the \emph{tetraquark} initial state for different values of $K$ ranging between $0.1J$ to $0.9J$ and for different values of $h$ ranging between $1J$ and $10J$ for a fixed $m=5J$. We plot these observables as a function of $(m-h)$ and see an expected peak when $h=m$. As we go further away from resonance, we see a sharp decay of the long-time \emph{tetraquark} and the $3$-meson numbers signifying the breakdown of the bound state physics. These curves get wider as $K/J$ increases. 

\section{Exact Diagonalization Spectrum}

\begin{figure}
    \centering
    \includegraphics[width=0.7\linewidth]{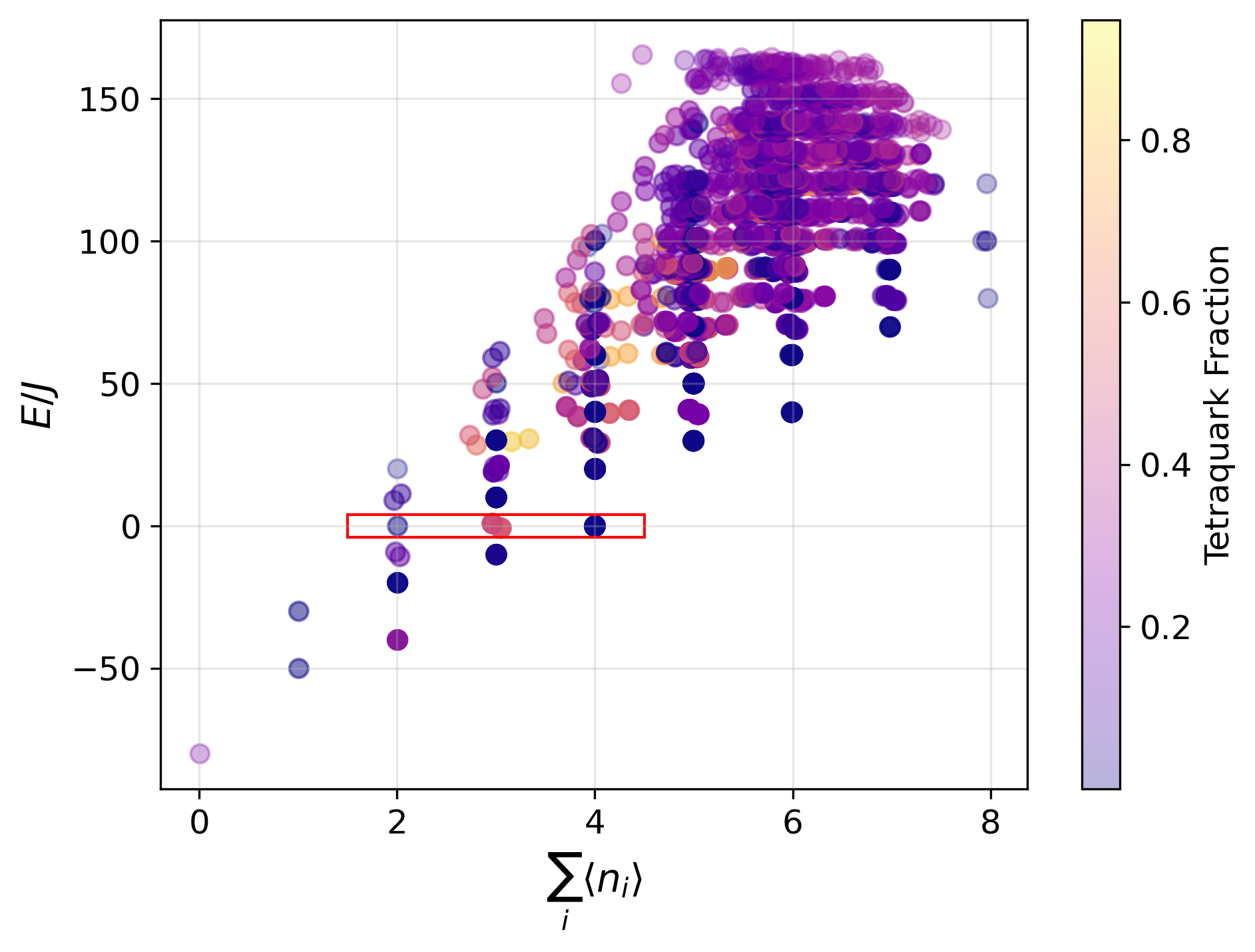}
    \caption{Full spectrum of the spin model defined in Eq. 2 of the main text for  $K=0.8J$, $h=m=10J$, and $L=12$. The states within the red rectangle are the relevant $3$-meson, \emph{tetraquark} and $1$-meson continuum states of interest. The energies of all the states been shifted such that the energies of the states of interest mentioned above are around $0$ for convenience.}
    \label{fig:full_spectrum}
\end{figure}

We have performed exact diagonalization (ED) calculations to complement the dynamics presented in the main text, and support our claims that for large $K$, we observe regular attractive bound states and for all $K$, repulsively bound states consisting of four particles. We diagonalize the gauge theory Hamiltonian (Eq.~2 in the main text) to compute the exact energy spectrum for a $L=12$ system. In Fig.~\ref{fig:full_spectrum}, we show the full energy spectrum with the horizontal coordinate indicating the total particle number in each eigenstate, for $K = 0.8J$ and $h=m=10J$. Increasing $h=m$ increases the separation between states in bands separated by energy $h$, corresponding to different number of particles or different sizes of mesons. In this limit, we can ignore coupling between these bands and focus on a single band within the full spectrum that contains states with a single $3$-meson, single \emph{tetraquark}, and two separated $1$-mesons. These are in the red rectangle shown in Fig.~\ref{fig:full_spectrum}. 

\begin{figure}
    \centering
    \includegraphics[width=0.8\linewidth]{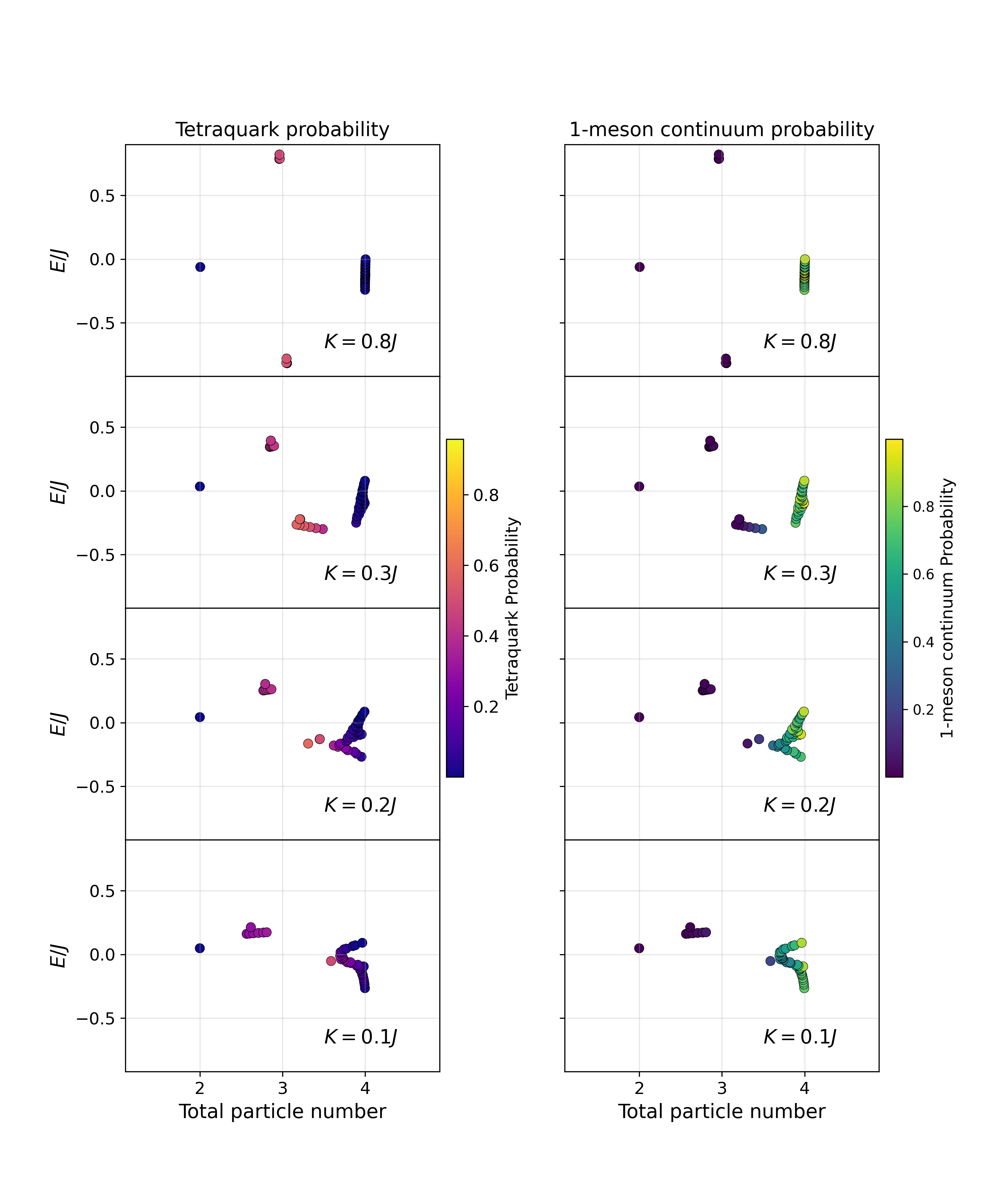}
    \caption{Full ED spectrum of the spin model defined in Eq. 2 of the main text for $L=12$ zoomed in to the states of interest. We see that as we decrease $K/J$, we move from having both regular attractive and repulsive bound states to exclusively repulsive bound states.}
    \label{fig:zoomed_ED}
\end{figure}

\begin{figure}
    \centering
    \includegraphics[width=0.8\linewidth]{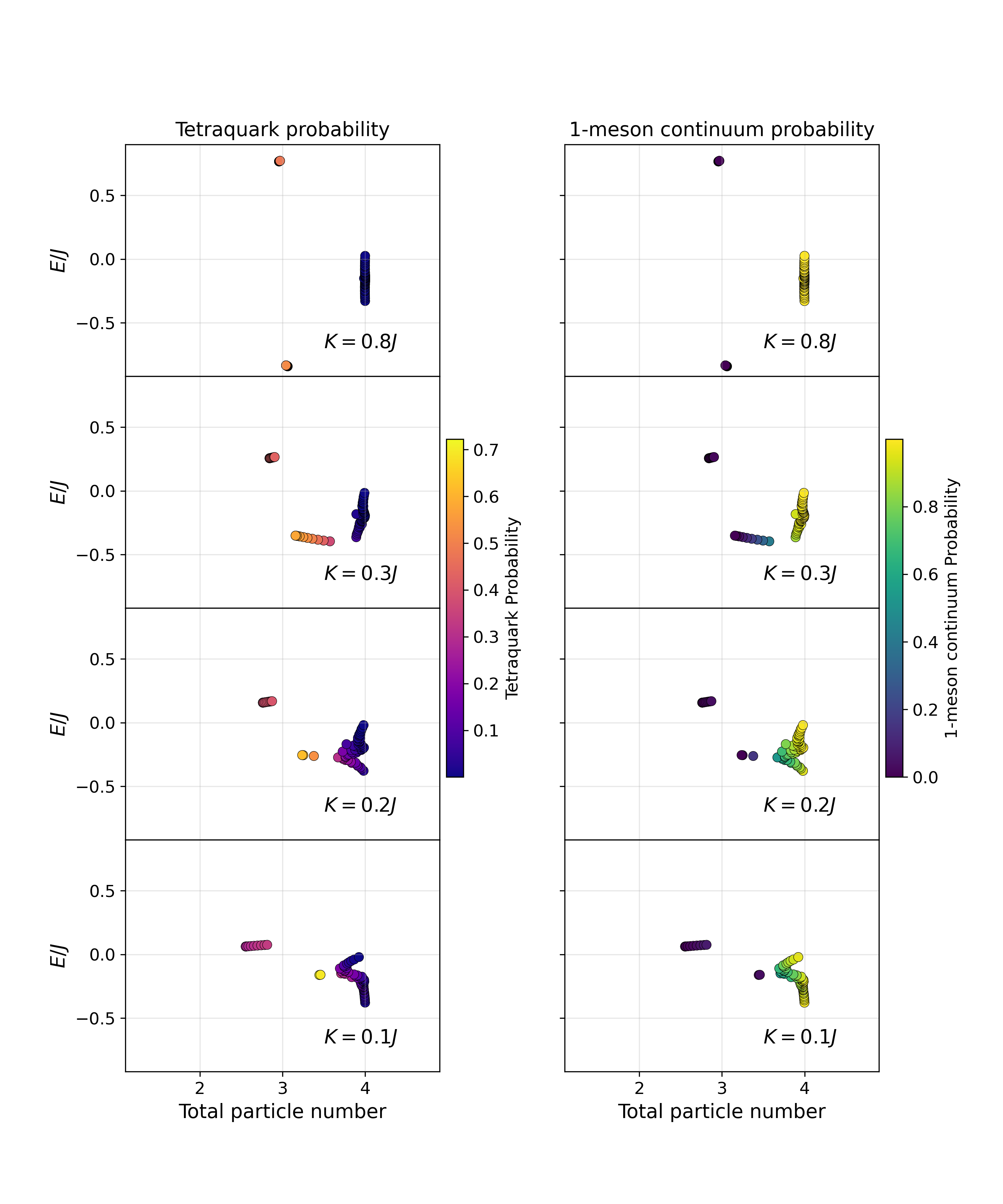}
    \caption{Full ED spectrum of the effective model defined in Eq. 4 of the main text for $L=12$. We see that the effective model captures all the states that the full ED of the spin model captures except some boundary states that are not relevant to the dynamics.}
    \label{fig:full_effective}
\end{figure}

 Fig.~\ref{fig:zoomed_ED} plots the spectrum over the range of this red rectangle for different values of $K/J$. We additionally compute the full spectrum of the effective model Hamiltonian shown in Eq. 4 of the main text. The full ED spectrum of the spin model in Eq. 2 is calculated for an $L=12$ site system. We also show the full ED spectrum of the effective model in Eq. 4 for an $L=12$ site system in Fig.~\ref{fig:full_effective}. We note that both  spectra look similar and the effective model captures all the states of the spin model except some states with total particle number $2$. These are boundary states of the form $\ket{\uparrow \downarrow \downarrow\downarrow\downarrow\downarrow\uparrow\uparrow}$ and $\ket{\uparrow\uparrow \downarrow \downarrow\downarrow\downarrow\downarrow\uparrow}$. These states correspond to a $3$-meson that wraps around the boundary. In our dynamical simulations, we consider open boundary conditions and thus such states are never produced. We neglect these states in the effective model since they have no effect on the physics in the large enough system sizes that we consider in our numerical simulations.

To analyze the ED spectrum in Fig.~\ref{fig:zoomed_ED}, we first observe that for $K=0.8J$ (the large $K/J$ limit)  states with large \emph{tetraquark} probability and small $1$-meson probability sit well above or below the continuum with a significant gap. In this limit, we have both the regular attractive bound states and repulsive bound states. As we lower $K/J$, we see that the gap reduces and the lower energy states with large \emph{tetraquark} probability gradually merge with the continuum. However, the higher energy repulsive bound states continue to have some energy gap relative to the continuum. The states that have high \emph{tetraquark} probability also have a high $3$-meson probability and thus particle number $\sum \langle n_i \rangle \sim 3$ (it would be exactly equal to $3$ if they were in an equal superposition). We note that the essential physics is captured already for the small system size of $L=12$. The effective model spectrum in Fig.~\ref{fig:full_effective} has similar features, showing that it captures the physics of the full model well in the large $m=h$ limit. We additionally note that the bandwidth of the continuum in the spectrum is $\sim 0.4J = 8J^2/2h$ as predicted in the main text.

\bibliography{main_text}